\RequirePackage{rotating}
\documentclass[a4paper,fleqn,usenatbib]{mnras}
\usepackage{mathptmx}
\usepackage{morefloats}
\usepackage[T1]{fontenc}
\usepackage{ae,aecompl}

\usepackage{graphicx}	
\usepackage{amsmath}	
\usepackage{amssymb}	
\usepackage{multicol}
\usepackage{color}
\usepackage{subfig}
\usepackage{float}
\usepackage{rotating}

\title[The OmegaWhite survey]{The OmegaWhite survey for Short-Period Variable Stars III: Follow-up Photometric and Spectroscopic Observations }

\author[S. Macfarlane et al.]  
{S.A~Macfarlane$^{1,2}$,
  P.A~Woudt$^2$,
  P.J~Groot$^1$,
  G.~Ramsay$^3$,  
  R.~Toma$^3$,
  \newauthor
  M.~Motsoaledi$^{2,4}$,
  L.A.~Crause$^4$,
  D.G.~Gilbank$^4$,
  D.~O'Donoghue$^4$\thanks{Deceased, 25 June 2015},
  \newauthor
  S.B.~Potter$^4$,
  A.A.~Sickafoose$^4$,
  C.~van Gend$^4$,
  H.L.~Worters$^4$
  \\
  $^1$Department of Astrophysics/IMAPP, 
      Radboud University, 
      P.O. Box 9010,
      6500 GL Nijmegen,
      The Netherlands \\
  $^2$Department of Astronomy, 
      University of Cape Town, 
      Private Bag X3, \\
      Rondebosch 7701, 
      South Africa \\      
  $^3$Armagh Observatory, 
      College Hill, 
      Armagh, 
      BT61 9DG,
      Northern Ireland\\
  $^4$South African Astronomical Observatory, 
      PO Box 9,
      7935 Observatory,
      South Africa \\
}

\date{Accepted XXX. Received YYY; in original form ZZZ}

\pubyear{2016}

\begin{document}
\label{firstpage}
\pagerange{\pageref{firstpage}--\pageref{lastpage}}
\maketitle

\begin{abstract}
We present photometric and spectroscopic follow-up observations of short-period variables discovered in the OmegaWhite survey:  a wide-field high-cadence $g$-band synoptic survey targeting the Galactic Plane. We have used fast photometry on the SAAO 1.0-m and 1.9-m telescopes to obtain light curves of 27 variables, and use these results to validate the period and amplitude estimates from the OmegaWhite processing pipeline.  Furthermore, 57 sources (44 unique, 13 also with new light curves) were selected for spectroscopic follow-up using either the SAAO 1.9-m telescope or the Southern African Large Telescope. We find many of these variables have spectra which are consistent with being $\delta$\,Scuti type pulsating stars. At higher amplitudes, we detect four possible pulsating white dwarf/subdwarf sources and an eclipsing cataclysmic variable. Due to their rarity, these targets are ideal candidates for detailed follow-up studies. From spectroscopy, we confirm the symbiotic binary star nature of two variables identified as such in the SIMBAD database. We also report what could possibly be the first detection of the `Bump Cepheid' phenomena in a $\delta$\,Scuti star, with OW\,J175848.21--271653.7 showing a pronounced 22\% amplitude dip lasting 3 minutes during each pulsational cycle peak. However, the precise nature of this target is still uncertain as it exhibits the spectral features of a B-type star.
\end{abstract}



\section{Introduction}

High cadence photometric observations of stars can reveal intriguing features in the resulting light curves which would typically be missed in lower cadence data. Until a decade ago, wide field surveys with high cadence were not common. Due to a rapid increase in technological capabilities, it is only within the last few years that many short-period stellar systems have been discovered in wide-field and high-cadence surveys such as the Kepler mission \citep{Borucki2010}, the OGLE-IV Real-Time Transient Search \citep{Wyrzykowski2014}, the Rapid Temporal Survey \citep[RATS,][]{Ramsay2005,Barclay2011}, and the OmegaWhite (OW) survey \citep[OW; ][]{Macfarlane2015}. This paper presents the results of photometric and spectroscopic follow-up observations of variable candidates discovered in the OW survey. It allows further characterisation of the selected targets and provides a means to confirm the validity of the light curve properties determined during OW data processing. Furthermore, by correlating the spectroscopic identification with the object's location in colour-colour or period-amplitude-magnitude space, we can improve target selection for future follow-up observations.

The aim of the OW survey is to identify short-period variable systems that exhibit modulations with a period under an hour, such as pulsating subdwarf stars \citep[sdB, sdOB or sdO,][and see \citet{Woudt2006} for the first sdO pulsator discovery]{Heber2009}, pulsating white dwarf stars (DAV, DBV), $\delta$\,Scuti stars \citep[$\delta$\,Sct,][]{Breger2000}, or the rare and faint hydrogen-deficient ultracompact binary star systems (UCBs). These include eclipsing detached double-degenerate systems \citep{Kilic2011,Hermes2012,Kilic2014} and semi-detached mass-transferring AM\,CVn systems \citep[for a detailed review, see][]{Solheim2010}. These short-period sources are astrophysically interesting for a number of reasons. Pulsating stars are important for stellar astroseismology research and our understanding of stellar interiors. Additionally, we can use the frequency of the fundamental radial mode exhibited by high-amplitude $\delta$\,Sct stars (those with amplitudes $\gtrsim$ 0.3 mag) for precise distance determinations \citep[e.g.][]{Peterson1999, McNamara2007}. In addition to providing insight into late-stage binary evolution research, the shortest-period UCBs are predicted to be the strongest known emitters in the passband of the gravitational wave satellite observatory {\sl eLISA} \citep{Amaro-Seoane2013,Roelofs2007}. 

The primary goal of the OW survey is to increase and characterise the known population of UCBs. Spectroscopic observations provide immediate classification of the nature of objects through their chemical composition and spectral appearance. For example, AM\,CVn binaries exhibit helium lines but no hydrogen, while pulsating sdB stars show deep hydrogen absorption lines, and accreting binaries with main sequence secondaries will show hydrogen emission lines. $\delta$\,Sct stars typically exhibit Balmer absorption lines, with evidence of sodium and magnesium with strengths characteristic of late A-type, or F-type stars. 

The OW survey aims to cover an area of 400 square degrees through high-cadence optical observations along the Galactic Plane reaching a depth of $g$ $\sim$ 21.5 mag (10$\sigma$). Survey operations began in December 2011 with ESO semester 88, and so far 230 sq. degr. have been observed throughout ESO semesters 88 to 96 using OmegaCAM on the VLT Survey Telescope \citep[VST,][]{Capaccioli2011}. The OW field pointings have been chosen such that they overlap with the VST Photometric H$\alpha$ Survey of the Southern Galactic Plane \citep[VPHAS+,][]{Drew2014} and the Galactic Bulge Survey \citep[GBS,][]{Jonker2011} in order to obtain broad-band colours and photometric zeropoints. The results from the analysis of the initial 26 sq. degr. are presented in the first OW paper \citep[hereafter Paper$\,$I;][]{Macfarlane2015}, which also details the processing pipeline the survey has adopted to optimally detect these short-period systems. Furthermore, detailed population studies of sources from the first 134 sq. degr. are presented in \citet[][hereafter Paper\,II]{Toma2016}.   

Photometric follow-up observations are important in determining the validity of the period and amplitude determined during the OW processing pipeline. With this goal in mind, this paper examines the photometric observations of 27 variable candidates selected using the OW pipeline. Furthermore we present the spectroscopic identification of 57 OW targets and analyse their period, amplitude and colour properties in order to optimise future follow-up target selection.

\section{Photometric Follow-up Observations}

Photometric observations of 27 targets were obtained using the Sutherland High-speed Optical Cameras \citep[SHOC,][]{Coppejans2013} on the 1.0-m and 1.9-m South African Astronomical Observatory (SAAO) telescopes, located at the Sutherland Observatory in South Africa. Their light curve and colour properties are shown in Table~\ref{TAB:photproperties}. These targets were selected as they appear relatively blue according to their VPHAS+ colour information, and they exhibit short-period variations in their OW light curves (with periods $<$ 70 mins) with low false alarm probabilities ($\log_{10}$(FAP) $<$ --2.5, see Section~\ref{SEC:Photcomp}).

\begin{table*}
\centering
\caption{Properties of targets for photometric follow-up observations: star ID; RA and Dec; OW Lomb Scargle period (P$_{OW}$) and false alarm probability ($\log_{10}$(FAP)) of the highest peak in the periodogram, OW calibrated $g$-band magnitude(OW$_{g}$), amplitude of OW light curve ($A_{OW}$), VPHAS+ colour indices $g - r$, $u - g$, $r - i$, and $r - H\alpha$.}
\resizebox{\textwidth}{!}{
\begin{tabular}{lrrrcrrrrrrl}
\hline
Star ID                & RA (J2000)  & DEC (J2000) & P$_{OW}$ & log$_{10}$(FAP) & g$_{OW}$  &$A_{OW}$ & $u - g$ & $g - r$  & $r - i$ & $r - H\alpha$  & comments\\
                       & (hh:mm:ss)  &   (\degr:\arcmin:\arcsec)  & (min)    &                 &   (mag)  &   (mag)           & (mag)   & (mag)   &(mag)   &  (mag)  & \\
\hline 
OW\,J073919.72--300922.7	&	07:39:19.72	&	--30:09:22.7			&	17.4	&	--2.83	&	16.26	&	0.005	&	0.33	&	0.51	&	0.31	&	0.15	&		\\
OW\,J074106.40--293325.5$^{a,b}$	&	07:41:06.40	&	--29:33:25.5	&	38.9	&	--4.89	&	15.75	&	0.009	&	0.28	&	0.51	&	0.26	&	0.12	&	$\delta$\,Sct (F3V/F4V)	\\
OW\,J074447.99--262155.6$^{b}$	&	07:44:47.99	&	--26:21:55.6	&	25.2	&	--3.82	&	16.30	&	0.018	&		&		&		&		&	$\delta$\,Sct (F3V/F4V)	\\
OW\,J075114.46--270311.2$^{b}$	&	07:51:14.46	&	--27:03:11.2	&	37.6	&	--3.72	&	15.69	&	0.011	&		&		&	0.46	&	0.20	&	$\delta$\,Sct (F6V/F7V)	\\
OW\,J075121.36--270916.4$^{b}$	&	07:51:21.36	&	--27:09:16.4	&	40.8	&	--4.98	&	14.63	&	0.021	&		&		&		&		&	$\delta$\,Sct (F2V)	\\
OW\,J075359.78--275732.6$^{a,b}$	&	07:53:59.78	&	--27:57:32.6	&	26.1	&	--5.28	&	13.98	&	0.020	&	0.37	&	0.33	&	0.20	&	0.09	&	$\delta$\,Sct (F3V/F4V)	\\
OW\,J075532.58--280854.3$^{a,b}$	&	07:55:32.58	&	--28:08:54.3	&	21.4	&	--4.17	&	16.70	&	0.018	&	0.50	&	0.52	&	0.34	&	0.13	&	$\delta$\,Sct (F5V)	\\
OW\,J075837.37--292904.2$^{b}$	&	07:58:37.37	&	--29:29:04.2	&	24.7	&	--4.24	&	16.30	&	0.011	&		&		&		&		&	$\delta$\,Sct (F3V/F4V)	\\
OW\,J075952.17--292817.6$^{b}$	&	07:59:52.17	&	--29:28:17.6	&	38.1	&	--4.37	&	15.05	&	0.012	&	0.20	&	0.48	&	0.28	&	0.28	&	$\delta$\,Sct (F5V)	\\
OW\,J080205.41--295804.6$^{b}$	&	08:02:05.41	&	--29:58:04.6	&	37.2	&	--4.13	&	16.76	&	0.016	&	0.57	&	0.81	&	0.46	&	0.29	&	$\delta$\,Sct (F3V/F4V)	\\
OW\,J080257.23--292136.0$^{b}$	&	08:02:57.23	&	--29:21:36.0	&	65.8	&	--3.62	&	14.71	&	0.014	&	0.02	&	0.76	&		&		&	$\delta$\,Sct (F6V/F7V)	\\
OW\,J080418.75--293740.2$^{b}$	&	08:04:18.75	&	--29:37:40.2	&	30.3	&	--3.37	&	14.17	&	0.005	&	0.14	&	0.42	&	0.26	&	0.10	&	$\delta$\,Sct (F5V)	\\
OW\,J080729.17--293417.1			&	08:07:29.17	&	--29:34:17.1	&	61.5	&	--5.19	&	17.00	&	0.031	&	0.22	&	0.68	&	0.36	&	0.16	&		\\
OW\,J080916.72--292158.0$^{a,b}$	&	08:09:16.72	&	--29:21:58.0	&	31.2	&	--5.78	&	15.30	&	0.027	&	0.33	&	0.45	&	0.27	&	0.19	&	$\delta$\,Sct (F5V)	\\
OW\,J174308.60--255102.6			&	17:43:08.60	&	--25:51:02.6	&	43.3	&	--3.57	&	14.05	&	0.008	&	0.67	&	0.84	&	0.72	&	0.22	&		\\
OW\,J174436.19--253349.6			&	17:44:36.19	&	--25:33:49.6	&	49.1	&	--3.21	&	16.22	&	0.027	&		&		&		&		&		\\
OW\,J175122.15--260916.7			&	17:51:22.15	&	--26:09:16.7	&	24.5	&	--4.39	&	14.44	&	0.005	&	0.38	&	0.63	&	0.34	&		&		\\
OW\,J175210.09--262535.0			&	17:52:10.09	&	--26:25:35.0	&	34.5	&	--3.75	&	15.61	&	0.008	&	0.74	&		&	0.60	&	0.18	&		\\
OW\,J175210.34--262048.7			&	17:52:10.34	&	--26:20:48.7	&	26.7	&	--5.02	&	14.69	&	0.018	&	0.48	&	0.69	&	0.39	&	0.05	&		\\
OW\,J175509.75--260225.5			&	17:55:09.75	&	--26:02:25.5	&	22.7	&	--4.81	&	15.31	&	0.015	&	0.56	&	0.70	&	0.57	&	0.20	&		\\
OW\,J175720.15--270100.5			&	17:57:20.15	&	--27:01:00.5	&	23.1	&	--4.31	&	15.75	&	0.024	&	0.72	&	0.87	&	0.54	&	0.17	&		\\
OW\,J175848.21--271653.7$^{b}$	&	17:58:48.21	&	--27:16:53.7	&	32.5	&	--3.49	&	15.75	&	0.254	&		&	0.40	&	0.27	&	0.12	&	$\delta$\,Sct with dips	\\
OW\,J175902.86--270025.1			&	17:59:02.86	&	--27:00:25.1	&	33.8	&	--3.19	&	15.76	&	0.017	&	0.56	&	0.82	&	0.53	&	0.20	&		\\
OW\,J175954.94--264752.1			&	17:59:54.94	&	--26:47:52.1	&	28.3	&	--3.48	&	16.36	&	0.014	&	0.86	&	1.08	&		&	0.30	&		\\
OW\,J180259.04--271204.3			&	18:02:59.04	&	--27:12:04.3	&	23.7	&	--3.98	&	16.58	&	0.024	&	1.28	&	1.27	&	0.74	&	0.19	&		\\
OW\,J180735.09--250943.8			&	18:07:35.09	&	--25:09:43.8	&	22.1	&	--3.44	&	16.33	&	0.012	&	0.96	&	1.08	&	0.69	&	0.26	&		\\
OW\,J181847.29--251231.9			&	18:18:47.29	&	--25:12:31.9	&	7.8	&	--4.89	&	15.83	&	0.028	&		&		&		&		&		\\
\hline
\multicolumn{12}{p{17.0cm}}{$^{a}$Observed spectroscopically with the RSS on SALT. P$_{OW}$ is equivalent to the P$_{LS}$ notation used in Paper$\,$I.}\\
\multicolumn{12}{p{17.0cm}}{$^{b}$Also a spectroscopic follow-up target, see Table~\ref{TAB:specproperties}.}\\
\end{tabular}}
\label{TAB:photproperties}
\end{table*}

The SHOC cameras are two identical high-time resolution cameras, each consisting of an Andor iXon X3 888 UVB camera with a 1024 $\times$ 1024 pixel, back-illuminated CCD, capable of accurate GPS-triggered optical timing observations with minimal dead-time. The cameras are mounted at the Cassegrain focus of the 1.0-m and 1.9-m telescopes, at focal ratios of $f$/16 and $f$/18 respectively. The field of view is 2.85 $\times$ 2.85 arcmin$^{2}$ for the 1.0-m, and 1.29 $\times$ 1.29 arcmin$^{2}$ for the 1.9-m telescope, respectively. 

White-light unfiltered observations were taken during four separate observing runs in April 2014, September 2014, December 2014 -- January 2015, and June 2015, as shown in the photometric observing log in Table~\ref{TAB:obslog_phot}. All targets were observed in frame-transfer mode, with a 1 MHz readout speed (conventional mode) and with a pre-amplifier gain of 2.4. Exposure times (as detailed in Table~\ref{TAB:obslog_phot}) were between 2 and 30 seconds.

The raw datacube images produced by SHOC  were reduced and light curves extracted using standard \textsc{IRAF} \citep{Tody1986,Tody1993} routines. Firstly, the datacubes were partitioned into individual science frames, which were then bias-subtracted and flat-fielded. Next, the specific Julian date and appropriate header information was assigned to each subsequent image. Using the technique of differential photometry, aperture photometry was applied to the target and a number of bright neighbouring stars in order to extract background-subtracted light curves for these sources. The target light curve is then divided by the average light curve of the reference stars in order to compensate for atmospheric effects during the observing runs.

The Fourier transform properties of the targets were determined by applying the same procedure detailed in Paper$\,$I, i.e.  we applied the \textsc{VARTOOLS} suite of software \citep[][]{Hartman2008} to our processed SHOC light curves, and were able to determine a range of photometric variability parameters, including the Lomb Scargle period (P$_{SHOC}$) and amplitude ($A_{SHOC}$). 

\subsection{Comparison between OW light curves and follow-up photometry}
\label{SEC:Photcomp}

The individual follow-up light curves of the 27 stars in our photometric sample are
presented in Figure~\ref{FIG_lcft}. Here we focus on the
properties of the sample and compare them with the light curves processed through the OW survey pipeline.

Firstly, we see that most of the targets are indeed variable in their SHOC light curves,  as 
should be the case since these were selected from OW as high-confidence candidates, having a false-alarm probability $\log_{10}$(FAP) $< -2.5$. Most of the variations seen in the SHOC-observed light curves in Figure~\ref{FIG_lcft} resemble the variations exhibited by their respective OW light curve, also shown in Figure~\ref{FIG_lcft}.

\begin{figure*}
\centering
\includegraphics[width=\linewidth]{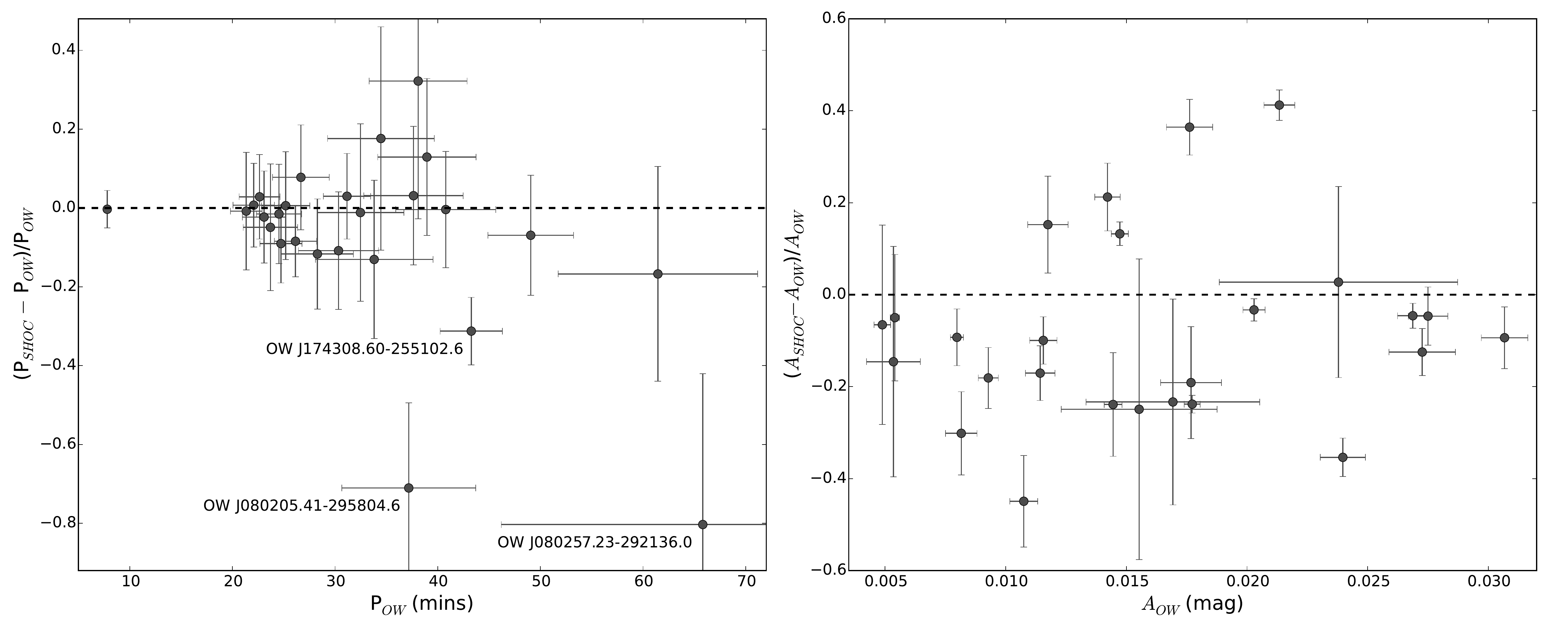}
\caption{Comparison of derived periods (P$_{SHOC}$) and amplitudes ($A_{SHOC}$) from the 27 light curves of SHOC-observed follow-up targets, to the respective OW derived periods (P$_{OW}$) and amplitudes ($A_{OW}$). A dashed line in each of the panels indicate the ideal case where either P$_{SHOC}$ = P$_{OW}$ (left panel), or $A_{SHOC}$ = $A_{OW}$ (right panel). Uncertainties in the values are indicated by horizontal and vertical bars. The target OW\,J073919.72--300922.7 is not shown in the left panel as its period ratio is too high ((P$_{SHOC}$ -- P$_{OW}$)/P$_{OW} \approx$ 2.6)}
\label{Fig_comp}
\end{figure*}

In Figure~\ref{Fig_comp} and Table \ref{TAB:validity} we present the derived SHOC periods (P$_{SHOC}$) and amplitudes ($A_{SHOC}$) together with those derived from OW data (P$_{OW}$ and $A_{OW}$ resp.). Figure~\ref{Fig_comp} shows that, for the majority of the targets, there is a strong correspondence between the periods derived from SHOC data and the periods derived from OW data. The median level of the deviation of the SHOC periods with respect to OW for all 27 targets is 8\%, with an RMS-scatter of 55\% which can be greatly improved if we do not consider the four targets with large deviations caused by factors such as poor observing conditions. For the amplitudes, the median level of the deviation from the OW amplitudes is 15\% with a RMS-scatter of 21\%. These relatively large amplitude deviations could be due to the fact that while OW observations were taken in Sloan $g$ band, all SHOC observations were observed in white light (with no filter). We therefore have to remain cautious when comparing these amplitudes.

\begin{table*} 
\centering
\caption{OW variable candidates selected from our SHOC photometric follow-up campaign, with OW Lomb Scargle period (P$_{OW}$) and associated amplitude ($A_{OW}$), and the Lomb Scargle period for the respective SHOC-observed light curves (P$_{SHOC}$) with associated amplitude ($A_{SHOC}$). }
\begin{tabular}{crrrr}
\hline
 Star ID 		& P$_{OW}$ &  $A_{OW}$	&P$_{SHOC}$ & $A_{SHOC}$    \\
         & (min)    &(mag)		& (min)	&  (mag)  \\ \hline  
OW\,J181847.29--251231.9	&	7.8	$\pm$	0.2	&	0.028	$\pm$	0.001	&	7.8	$\pm$	0.3	&	0.026	$\pm$	0.002	\\
OW\,J073919.72--300922.7$^{*}$	&	17.4	$\pm$	1.0	&	0.005	$\pm$	0.001	&	62.4	$\pm$	8.6	&	0.005	$\pm$	0.001	\\
OW\,J075532.58--280854.3	&	21.3	$\pm$	1.5	&	0.018	$\pm$	0.001	&	21.2	$\pm$	2.8	&	0.014	$\pm$	0.002	\\
OW\,J180735.09--250943.8	&	22.1	$\pm$	2.0	&	0.012	$\pm$	0.001	&	22.2	$\pm$	1.2	&	0.010	$\pm$	0.001	\\
OW\,J175509.75--260225.5	&	22.6	$\pm$	2.0	&	0.015	$\pm$	0.001	&	23.3	$\pm$	1.4	&	0.017	$\pm$	0.001	\\
OW\,J175720.15--270100.5	&	23.1	$\pm$	2.1	&	0.024	$\pm$	0.001	&	22.6	$\pm$	1.7	&	0.015	$\pm$	0.001	\\
OW\,J180259.04--271204.3	&	23.7	$\pm$	2.7	&	0.024	$\pm$	0.005	&	22.5	$\pm$	2.7	&	0.024	$\pm$	0.001	\\
OW\,J175122.15--260916.7	&	24.5	$\pm$	2.2	&	0.005	$\pm$	0.001	&	24.2	$\pm$	2.2	&	0.005	$\pm$	0.001	\\
OW\,J075837.37--292904.2	&	24.7	$\pm$	2.0	&	0.011	$\pm$	0.001	&	22.5	$\pm$	1.4	&	0.009	$\pm$	0.000	\\
OW\,J074447.99--262155.6	&	25.2	$\pm$	2.3	&	0.018	$\pm$	0.001	&	25.3	$\pm$	2.5	&	0.024	$\pm$	0.001	\\
OW\,J075359.78--275732.6	&	26.1	$\pm$	2.1	&	0.020	$\pm$	0.001	&	23.9	$\pm$	1.1	&	0.020	$\pm$	0.001	\\
OW\,J175210.34--262048.7	&	26.7	$\pm$	2.8	&	0.018	$\pm$	0.001	&	28.7	$\pm$	2.2	&	0.014	$\pm$	0.001	\\
OW\,J175954.94--264752.1	&	28.3	$\pm$	3.5	&	0.014	$\pm$	0.001	&	25.0	$\pm$	1.8	&	0.017	$\pm$	0.001	\\
OW\,J080418.75--293740.2	&	30.3	$\pm$	3.9	&	0.005	$\pm$	0.001	&	27.1	$\pm$	2.3	&	0.005	$\pm$	0.001	\\
OW\,J080916.72--292158.0	&	31.2	$\pm$	2.3	&	0.027	$\pm$	0.001	&	32.1	$\pm$	2.5	&	0.026	$\pm$	0.001	\\
OW\,J175848.21--271653.7	&	32.5	$\pm$	4.2	&	0.254	$\pm$	0.052	&	32.1	$\pm$	6.0	&	0.227	$\pm$	0.002	\\
OW\,J175902.86--270025.1	&	33.8	$\pm$	5.7	&	0.017	$\pm$	0.004	&	29.4	$\pm$	3.6	&	0.013	$\pm$	0.001	\\
OW\,J175210.09--262535.0	&	34.5	$\pm$	5.2	&	0.008	$\pm$	0.001	&	40.5	$\pm$	8.2	&	0.007	$\pm$	0.001	\\
OW\,J080205.41--295804.6$^{*}$	&	37.2	$\pm$	6.5	&	0.016	$\pm$	0.003	&	10.8	$\pm$	0.6	&	0.012	$\pm$	0.004	\\
OW\,J075114.46--270311.2	&	37.6	$\pm$	4.8	&	0.011	$\pm$	0.001	&	38.8	$\pm$	4.5	&	0.006	$\pm$	0.001	\\
OW\,J075952.17--292817.6	&	38.1	$\pm$	4.8	&	0.012	$\pm$	0.001	&	50.4	$\pm$	12.3	&	0.014	$\pm$	0.001	\\
OW\,J074106.40--293325.5	&	38.9	$\pm$	4.8	&	0.009	$\pm$	0.001	&	44.0	$\pm$	6.1	&	0.008	$\pm$	0.001	\\
OW\,J075121.36--270916.4	&	40.8	$\pm$	4.9	&	0.021	$\pm$	0.001	&	40.6	$\pm$	3.5	&	0.030	$\pm$	0.001	\\
OW\,J174308.60--255102.6$^{*}$	&	43.3	$\pm$	3.0	&	0.008	$\pm$	0.001	&	29.8	$\pm$	1.9	&	0.006	$\pm$	0.001	\\
OW\,J174436.19--253349.6	&	49.1	$\pm$	4.2	&	0.027	$\pm$	0.001	&	45.7	$\pm$	6.2	&	0.024	$\pm$	0.001	\\
OW\,J080729.17--293417.1	&	61.4	$\pm$	9.7	&	0.031	$\pm$	0.001	&	51.2	$\pm$	13.5	&	0.028	$\pm$	0.002	\\
OW\,J080257.23--292136.0$^{*}$	&	65.8	$\pm$	19.6	&	0.014	$\pm$	0.001	&	13.0	$\pm$	0.5	&	0.011	$\pm$	0.002	\\
\hline
\multicolumn{5}{p{10.0cm}}{$^{*}$ Large period differences explained in Section~\ref{SEC:Photcomp}}\\
\end{tabular}
\label{TAB:validity}
\end{table*}

Although the majority of the targets exhibit variations in their SHOC light curves that are similar to the variations seen in their respective OW light curve, a number of the SHOC light curves show unexpected behaviour, which is reflected in the large error bars and deviations shown in Figure~\ref{Fig_comp} and Table \ref{TAB:validity}. In particular, the SHOC and OW periods for four of the targets are significantly different. These targets are OW$\,$J073919.72--300922.7, OW$\,$J080205.41--295804.6, OW$\,$J174308.60--255102.6, and OW$\,$J080257.23--292136.0 (hereafter OW$\,$J0739--3009, OW$\,$J0802--2958, OW$\,$J1743--2551, and OW$\,$J0802--2921). If we do not consider these four targets, the median level of the deviation of the SHOC periods with respect to OW for the 23 targets becomes 5\%, with an RMS-scatter of 10\%. \\

These differences can be due to:
\begin{itemize}
\item Poor photometry for a portion of the SHOC light curve. Poor atmospheric conditions during observing (such as an increasing airmass or heavy clouds), can create multiple peaks in the Discrete Fourier Transform (DFT), confusing the peak period. For example, multiple peaks are shown in the DFT of OW\,J0802--2921, and although the peak OW frequency can be identified, it is not the dominant frequency (as shown in Figure~\ref{FIG_lcft}, where we compare the peak frequencies detected in the DFT of the OW light curves, to that of the SHOC light curves.)
\item Non-periodic variability behaviour in the OW light curve. In some of OW light curves, the period and amplitude varies over the duration of the observation. Although this behaviour can be seen in their respective SHOC light curves, this leads to greater uncertainties in the period and amplitude ratios.
\item Large gap in the OW light curve (due to VST technical issues or poor weather) decreasing the OW period accuracy. In the case of OW\,J1743--2551, the SHOC light curve appears more complicated than its OW counterpart, as indicated by the multiple DFT peaks in Figure~\ref{FIG_lcft}.
\item Marginal selection in the OW data. The probability that the OW period determined for OW\,J0739--3009 is due to noise is relatively high ($\log_{10}$(FAP) is high, i.e. not a strong outlier from the median).
\item Limited signal-to-noise ratios. Sources such as OW\,J0802--2958 were observed using SHOC during periods of very poor seeing on the relatively smaller SAAO 1.0-m telescope, greatly increasing the observed noise. In the light curve of OW\,J0802--2958, this created multiple DFT peaks from the SHOC light curve, the strongest peak not in agreement with that of OW.
\item Long-timescale changes in the light curves. The characteristics of some sources is known to vary over longer timescales.  For instance, CVs are known to exhibit state changes, where the amplitude and time scale of photometric variability may drastically change over time. 
\item Phasing phenomena. Pulsating sources are known for phenomena such as beating, which could strongly affect the amplitude of variability.
\item Relatively short OW observation duration. Pulsators such as $\delta$\,Sct stars can show different behaviour in snapshot observations such as the 2 hr light curves derived using OW data \citep[for e.g. see Figure 7 in ][]{Ramsay2014}.  
\end{itemize}

\section{Spectroscopic Follow-up Observations}

Identification spectra have been obtained using either the Spectrograph Upgrade -- Newly Improved Cassegrain spectrograph (SpUpNIC, Crause et al.,in prep) on the SAAO 1.9-m telescope, or the Robert Stobie Spectrograph \citep[RSS,][]{Kobulnicky2003} on the 10-m Southern African Large Telescope \citep[SALT][]{Buckley2006} or on both instruments. See Table~\ref{TAB:specproperties} for a list of the light curve and colour properties for all spectroscopic follow-up targets. Furthermore, the spectroscopic observing log is shown in Appendix Table~\ref{obslog_spec}. 

\begin{table*}
\centering
\caption{Properties of targets for spectroscopic follow-up observations: star ID; RA and Dec; OW Lomb Scargle period (P$_{OW}$) and false alarm probability ($\log_{10}$(FAP)) of the highest peak in the periodogram, OW calibrated $g$-band magnitude($g_{OW}$), amplitude of OW light curve ($A_{OW}$), VPHAS+ colour indices $g - r$, $u - g$, $r - i$, and $r - H\alpha$.}
\resizebox{\textwidth}{!}{
\begin{tabular}{lrrrcrrrrrrl}
\hline
Star ID                & RA (J2000)  & DEC (J2000) & P$_{OW}$ & log$_{10}$(FAP) & $g_{OW}$  &$A_{OW}$ & $u - g$ & $g - r$  & $r - i$ & $r - H\alpha$  & comments\\
                       & (hh:mm:ss)  &   (\degr:\arcmin:\arcsec)  & (min)    &                 &   (mag)  &   (mag)           & (mag)   & (mag)   &(mag)   &  (mag)  & \\
\hline 
OW\,J073649.27--295601.8			&	07:36:49.27	&	--29:56:01.8	&	95.8	&	--3.59	&	14.65	&	0.004	&--0.01	&	0.55	&	0.40	&	0.24	&	$\delta$\,Sct (F8V)	\\
OW\,J073823.16--303958.0			&	07:38:23.16	&	--30:39:58.0	&	9.0	&	--3.73	&	15.70	&	0.054	&		&		&		&		&	$\delta$\,Sct (F5V)	\\
OW\,J073909.96--300620.0			&	07:39:09.96	&	--30:06:20.0	&	89.6	&	--5.92	&	13.64	&	0.011	&	0.33	&	0.56	&	0.30	&	0.26	&	$\delta$\,Sct (F5V)	\\
OW\,J074059.42--242452.4			&	07:40:59.42	&	--24:24:52.4	&	40.0	&	--4.61	&	14.01	&	0.007	&		&		&		&		&	$\delta$\,Sct (F5V)	\\
OW\,J074101.18--291540.0			&	07:41:01.18	&	--29:15:40.0	&	69.8	&	--5.49	&	16.65	&	0.017	&		&		&		&		&	$\delta$\,Sct (F8V)	\\
OW\,J074106.07--294811.0$^{a}$	&	07:41:06.11	&	--29:48:11.2	&	22.6	&	--3.87	&	20.02	&	0.224	&--1.23	&	0.04	&	0.05	&		&	pulsating WD?	\\
OW\,J074106.40--293325.5$^{a,c}$	&	07:41:06.40	&	--29:33:25.5	&	38.9	&	--4.89	&	15.75	&	0.009	&	0.28	&	0.51	&	0.26	&	0.12	&	$\delta$\,Sct (F3V/F4V)	\\
OW\,J074127.67--243642.9			&	07:41:27.67	&	--24:36:42.9	&	37.4	&	--3.73	&	15.39	&	0.007	&	0.74	&	0.85	&	0.37	&	0.17	&	$\delta$\,Sct (F8V)	\\
OW\,J074216.77--241327.5			&	07:42:16.77	&	--24:13:27.5	&	30.5	&	--5.17	&	15.99	&	0.015	&		&		&		&		&	$\delta$\,Sct (F3V/F4V)	\\
OW\,J074447.99--262155.6$^{c}$	&	07:44:47.99	&	--26:21:55.6	&	25.2	&	--3.82	&	16.30	&	0.018	&		&		&		&		&	$\delta$\,Sct (F3V/F4V)	\\
OW\,J074513.22--261036.0$^{a}$	&	07:45:13.22	&	--26:10:36.0	&	38.2	&	--3.73	&	14.73	&	0.010	&	0.65	&	1.09	&	0.26	&	0.06	&	$\delta$\,Sct (F5V)	\\
OW\,J074515.73--260841.7			&	07:45:15.73	&	--26:08:41.7	&	74.4	&	--5.35	&	15.11	&	0.024	&	0.67	&	1.30	&	0.35	&		&	$\delta$\,Sct (F6V/F7V)	\\
OW\,J074533.75--263212.5			&	07:45:33.75	&	--26:32:12.5	&	73.9	&	--5.11	&	15.16	&	0.028	&	0.64	&	1.21	&	0.31	&	0.08	&	$\delta$\,Sct (F3V/F4V)	\\
OW\,J074551.57--242147.8			&	07:45:51.57	&	--24:21:47.8	&	38.7	&	--4.86	&	15.71	&	0.018	&		&		&		&		&	$\delta$\,Sct (F8V)	\\
OW\,J075049.73--270018.3			&	07:50:49.73	&	--27:00:18.3	&	42.3	&	--2.70	&	16.03	&	0.009	&		&		&	0.32	&	0.12	&	$\delta$\,Sct (F2V)	\\
OW\,J075114.46--270311.2$^{c}$	&	07:51:14.46	&	--27:03:11.2	&	37.6	&	--3.72	&	15.69	&	0.011	&		&		&	0.46	&	0.20	&	$\delta$\,Sct (F6V/F7V)	\\
OW\,J075121.36--270916.4$^{c}$	&	07:51:21.36	&	--27:09:16.4	&	40.8	&	--4.98	&	14.63	&	0.021	&		&		&		&		&	$\delta$\,Sct (F2V)	\\
OW\,J075305.78--301208.1$^{a}$	&	07:53:05.78	&	--30:12:08.1	&	102.2&	--3.23	&	16.92	&	0.030	&--0.74	&	0.46	&	0.28	&	0.24	&	early B-type$^{d}$	\\
OW\,J075359.78--275732.6$^{a,c}$	&	07:53:59.78	&	--27:57:32.6	&	26.1	&	--5.28	&	13.98	&	0.020	&	0.37	&	0.33	&	0.20	&	0.09	&	$\delta$\,Sct (F3V/F4V)	\\
OW\,J075436.40--304655.0$^{a}$	&	07:54:36.40	&	--30:46:55.0	&	11.6	&	--2.89	&	20.62	&	0.246	&--0.78	&	0.34	&		&		&	pulsating WD?	\\
OW\,J075527.61--314825.3$^{a}$	&	07:55:27.61	&	--31:48:25.3	&	38.6	&	--3.07	&	15.85	&	0.007	&--0.22	&	0.64	&	0.42	&		&	$\delta$\,Sct (F6V/F7V)	\\
OW\,J075531.59--315058.2			&	07:55:31.59	&	--31:50:58.2	&	9.6	&	--3.33	&	16.98	&	0.009	&	2.45	&	2.13	&	1.06	&	0.42	&	late G-type	\\
OW\,J075532.58--280854.3$^{a,c}$	&	07:55:32.58	&	--28:08:54.3	&	21.4	&	--4.17	&	16.70	&	0.018	&	0.50	&	0.52	&	0.34	&	0.13	&	$\delta$\,Sct (F5V)	\\
OW\,J075620.67--291605.4			&	07:56:20.67	&	--29:16:05.4	&	58.8	&	--4.31	&	16.96	&	0.040	&	0.53	&	0.96	&	0.48	&	0.18	&	$\delta$\,Sct F8V	\\
OW\,J075640.83--291107.9			&	07:56:40.83	&	--29:11:07.9	&	38.7	&	--3.86	&	15.83	&	0.009	&	0.37	&	0.53	&	0.28	&	0.08	&	$\delta$\,Sct (F5V)	\\
OW\,J075719.94--292955.5$^{a}$	&	07:57:19.94	&	--29:29:55.5	&	98.7	&	--3.42	&	20.24	&	0.436	&--0.93	&	0.14	&	0.02	&--0.23	&	subdwarf pulsator?	\\
OW\,J075837.37--292904.2$^{c}$	&	07:58:37.37	&	--29:29:04.2	&	24.7	&	--4.24	&	16.30	&	0.011	&		&		&		&		&	$\delta$\,Sct (F3V/F4V)	\\
OW\,J075918.25--313721.6			&	07:59:18.25	&	--31:37:21.6	&	40.0	&	--3.52	&	15.73	&	0.009	&	0.40	&	0.70	&	0.45	&		&	$\delta$\,Sct (F6V/F7V)	\\
OW\,J075952.17--292817.6$^{c}$	&	07:59:52.17	&	--29:28:17.6	&	38.1	&	--4.37	&	15.05	&	0.012	&	0.20	&	0.48	&	0.28	&	0.28	&	$\delta$\,Sct (F5V)	\\
OW\,J080146.04--294518.2			&	08:01:46.04	&	--29:45:18.2	&	64.0	&	--3.65	&	15.93	&	0.013	&	0.17	&	0.56	&	0.39	&	0.15	&	$\delta$\,Sct (F6V/F7V)	\\
OW\,J080154.61--281513.9			&	08:01:54.61	&	--28:15:13.9	&	49.4	&	--1.10	&	17.11	&	0.279	&		&		&		&		&	G5V	\\
OW\,J080202.79--302356.3			&	08:02:02.79	&	--30:23:56.3	&128.3	&	--3.32	&	15.81	&	0.018	&--0.04	&	0.72	&	0.39	&	0.28	&	G2V	\\
OW\,J080205.31--300331.4			&	08:02:05.31	&	--30:03:31.4	&	72.8	&	--4.33	&	14.54	&	0.216	&	0.18	&	0.55	&	0.32	&	0.22	&	HADs? (F6V/F7V)\\
OW\,J080205.41--295804.6$^{c}$	&	08:02:05.41	&	--29:58:04.6	&	37.2	&	--4.13	&	16.76	&	0.016	&	0.57	&	0.81	&	0.46	&	0.29	&	$\delta$\,Sct (F3V/F4V)	\\
OW\,J080242.91--314653.4			&	08:02:42.91	&	--31:46:53.4	&	40.1	&	--3.52	&	15.72	&	0.009	&	0.22	&		&		&	0.19	&	$\delta$\,Sct (F4V)	\\
OW\,J080257.23--292136.0$^{c}$	&	08:02:57.23	&	--29:21:36.0	&	65.8	&	--3.62	&	14.71	&	0.014	&	0.02	&	0.76	&		&		&	$\delta$\,Sct (F6V/F7V)	\\
OW\,J080311.83--291144.6$^{a}$	&	08:03:11.83	&	--29:11:44.6	&122.6	&	--4.40	&	19.74	&	0.502	&--1.77	&--0.20	&	0.04	&	0.91	&	ecl. CV	\\
OW\,J080313.04--320720.2			&	08:03:13.04	&	--32:07:20.2	&	22.8	&	--4.34	&	16.80	&	0.022	&	0.36	&	0.52	&	0.39	&	0.18	&	$\delta$\,Sct (F2V)	\\
OW\,J080323.23--314852.6			&	08:03:23.23	&	--31:48:52.6	&	37.9	&	--4.44	&	16.09	&	0.032	&	0.30	&	0.38	&	0.32	&	0.17	&	$\delta$\,Sct (F5V)	\\
OW\,J080418.75--293740.2$^{c}$	&	08:04:18.75	&	--29:37:40.2	&	30.3	&	--3.37	&	14.17	&	0.005	&	0.14	&	0.42	&	0.26	&	0.10	&	$\delta$\,Sct (F5V)	\\
OW\,J080517.37--310257.5			&	08:05:17.37	&	--31:02:57.5	&	20.7	&	--3.76	&	15.55	&	0.011	&	0.26	&	0.55	&	0.31	&	0.18	&	$\delta$\,Sct (F6V/F7V)	\\
OW\,J080527.94--275144.8			&	08:05:27.94	&	--27:51:44.8	&	48.0	&	--4.64	&	15.56	&	0.009	&	0.18	&	0.55	&	0.25	&	0.10	&	$\delta$\,Sct (F6V/F7V)	\\
OW\,J080531.16--274221.5			&	08:05:31.16	&	--27:42:21.5	&	73.8	&	--4.93	&	17.60	&	0.054	&	0.32	&	0.68	&		&		&	G1V	\\
OW\,J080626.33--302545.3			&	08:06:26.33	&	--30:25:45.3	&113.5	&	--3.50	&	15.68	&	0.018	&--0.03	&	0.85	&	0.41	&	0.26	&	G1V	\\
OW\,J080901.66--291652.2			&	08:09:01.66	&	--29:16:52.2	&	39.8	&	--3.62	&	16.24	&	0.016	&	0.21	&	0.47	&	0.27	&	0.12	&	$\delta$\,Sct (F6V/F7V)	\\
OW\,J080910.70--300744.1			&	08:09:10.70	&	--30:07:44.1	&108.2	&	--3.03	&	16.19	&	0.015	&--0.01	&	0.78	&	0.41	&	0.28	&	G1V	\\
OW\,J080911.11--300209.2			&	08:09:11.11	&	--30:02:09.2	&112.3	&	--3.63	&	15.96	&	0.020	&--0.04	&	0.68	&	0.32	&	0.23	&	$\delta$\,Sct (F5V)	\\
OW\,J080916.72--292158.0$^{a,c}$	&	08:09:16.72	&	--29:21:58.0	&	31.2	&	--5.78	&	15.30	&	0.027	&	0.33	&	0.45	&	0.27	&	0.19	&	$\delta$\,Sct (F5V)	\\
OW\,J081018.30--273749.1			&	08:10:18.30	&	--27:37:49.1	&	52.6	&	--4.94	&	15.49	&	0.038	&	0.19	&	0.54	&	0.30	&	0.16	&	$\delta$\,Sct F5V	\\
OW\,J081116.72--313821.1			&	08:11:16.72	&	--31:38:21.1	&	33.9	&	--3.64	&	14.83	&	0.007	&		&		&		&		&	$\delta$\,Sct (F5V)	\\
OW\,J081234.73--312409.8			&	08:12:34.73	&	--31:24:09.8	&	35.4	&	--4.12	&	15.02	&	0.011	&		&		&		&		&	$\delta$\,Sct (F6V/F7V)	\\
OW\,J081321.93--313537.7			&	08:13:21.93	&	--31:35:37.7	&	34.9	&	--3.50	&	14.86	&	0.006	&		&		&		&		&	$\delta$\,Sct (F6V/F7V)	\\
OW\,J082257.05--280051.0			&	08:22:57.05	&	--28:00:51.0	&	40.2	&	--3.64	&	15.52	&	0.011	&		&		&		&		&	$\delta$\,Sct (F8V)	\\
OW\,J172235.29--321403.6$^{b}$	&	17:22:35.27	&	--32:14:03.7	&	22.6	&	--4.96	&	15.75	&	0.006	&		&		&		&		&	PN?, Th 3-8	\\
OW\,J175848.21--271653.7$^{c}$	&	17:58:48.21	&	--27:16:53.7	&	32.5	&	--3.49	&	15.75	&	0.254	&		&	0.40	&	0.27	&	0.12	&	$\delta$\,Sct with dips	\\
OW\,J180441.23--270912.4$^{b}$	&	18:04:41.22	&	--27:09:12.4	&	22.8	&	--3.18	&	15.06	&	0.006	&		&		&		&		&	Sym, SS73\,122	\\
OW\,J181043.86--275750.0$^{b}$	&	18:10:43.86	&	--27:57:50.1	&	31.4	&	--3.62	&	14.69	&	0.008	&		&		&		&		&	Sym, Hen\,2--357	\\
\hline
\multicolumn{12}{p{17.0cm}}{$^{a}$Observed spectroscopically with the RSS on SALT. P$_{OW}$ is equivalent to the P$_{LS}$ notation used in Paper$\,$I.}\\
\multicolumn{12}{p{17.5cm}}{$^{b}$Identified in the SIMBAD database as either a planetary nebula candidate (PN?), or a symbiotic binary star system (Sym).}\\
\multicolumn{12}{p{16.0cm}}{$^{c}$also a photometric follow-up target, see Table~\ref{TAB:photproperties}.}\\
\multicolumn{12}{p{16.0cm}}{$^{d}$Classified through visual inspection of the relative intensities of the H, He I and Na I doublet absorption lines}\\
\end{tabular}}
\label{TAB:specproperties}
\end{table*}

Eleven targets have so far been observed using SALT between December 2014 and January 2016, with results for the first six targets presented in Paper$\,$I. All 11 sources were selected as they either exhibited light curve and colour characteristics most consistent with those of known UCBs, or they exhibited unusual features in their light curves. For example, three of the targets exhibited short-period variations (P$_{OW}<$ 40 mins) and very blue colours consistent with known AM\,CVn stars (OW\,J074106.1--294811.2, OW\,J075436.40--304655.0, and OW\,J075527.61--314825.3).  Whereas, two of the other targets, namely OW\,J080311.83--291144.6 and OW\,J075719.94--292955.5, exhibited relatively long periods  (P$_{OW}>$ 90 mins), but are some of the bluest OW sources, as derived from VPHAS+ photometry \citep{Drew2014}.

All SALT targets were observed using the RSS in long-slit mode, with a 1.5\arcsec\ slit width (see Table~\ref{obslog_spec} for the SALT observing log). Early observations were obtained using the pg0300 grating with a grating angle of 5 degrees, providing a wavelength range of $\sim$3200 \AA\ - 9000 \AA\ and a central dispersion of 1.52 \AA/pixel. In order to improve the spectral resolution for later observations, the pg0900 grating with a grating angle of 14.375 degrees was used, covering a smaller wavelength range of $\sim$3980 \AA\ - 6980 \AA\ with a dispersion of 0.49 \AA/pixel. All targets were reduced using either an Argon or a Xenon arc for wavelength calibration and the standard \textsc{IRAF} \citep{Tody1986,Tody1993} and \textsc{PYSALT} \citep{Crawford2014} routines. 

SpUpNIC longslit spectra were obtained for 47 targets, as shown in Table~\ref{obslog_spec}. These bright targets ($g <$ 18 mag) were selected for various reasons: a) the target had already been observed photometrically using SHOC, b) the target has SALT identification spectra to compare with, c) the target stands out in colour-colour space as a particularly blue or red source, or d) it exhibited a dominant short-period variation (P$_{OW}<$ 45 mins) with a low probability of being due to noise ($\log_{10}$(FAP) $<$ --4.5). Observations were made using grating 6 and grating 7, with varying slit widths and grating angles (see Table~\ref{obslog_spec} for details) to achieve approximate wavelength ranges of 3560 \AA\ - 6470 \AA (dispersion $=$ 1.39 \AA/pixel), or 3120 \AA\ - 8290 \AA\  (dispersion $=$ 2.85 \AA/pixel) respectively. CuAr arcs were taken for wavelength calibration purposes with either no order filter for grating 6 or the BG38 filter for grating 7, and all SpUpNIC spectra were processed and extracted using standard IRAF routines.

\subsection{Spectroscopic Classification}
\label{SEC:SpecResults}

\begin{figure*}
\centering
\includegraphics[width=\linewidth]{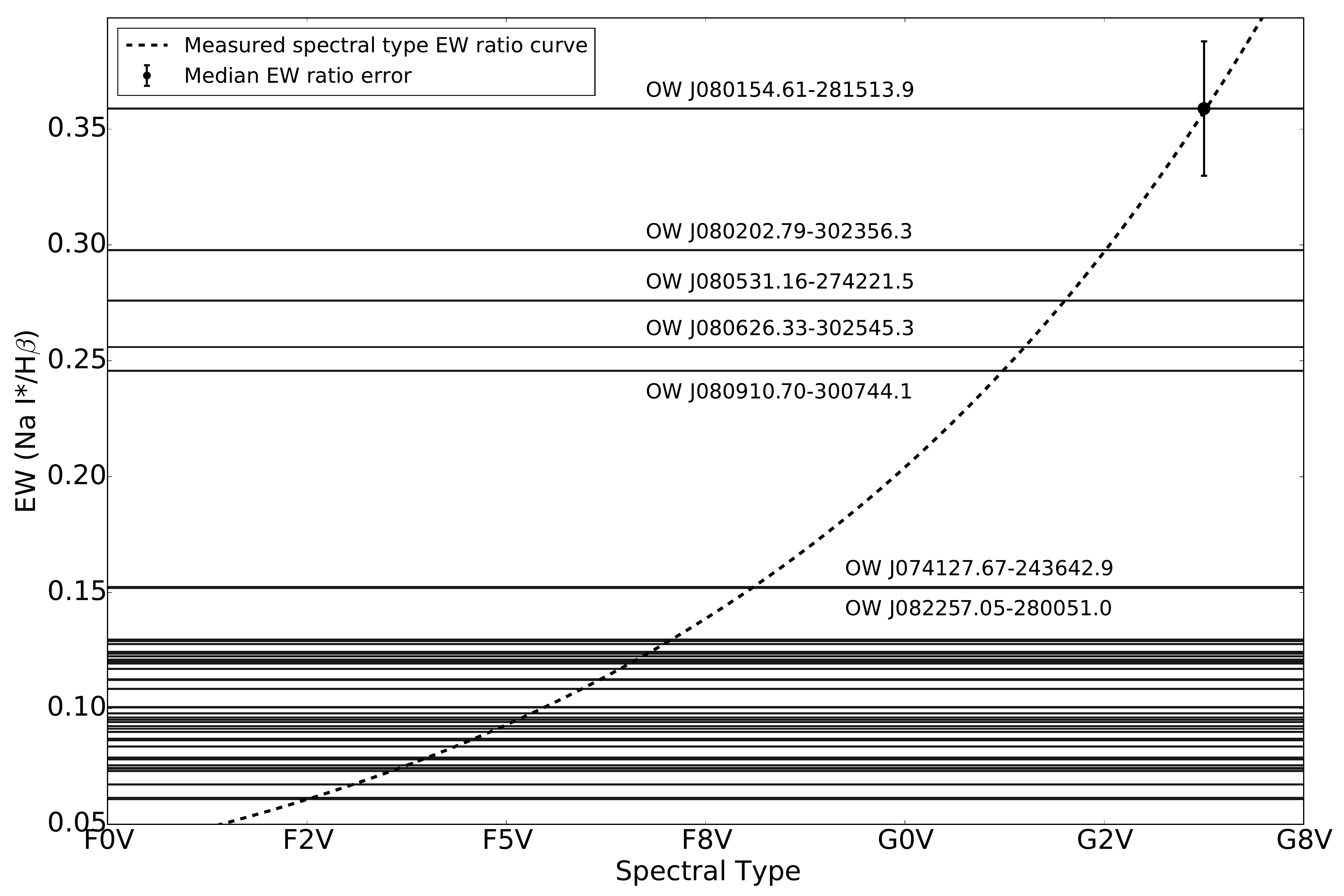}
\caption{Equivalent Width ratios of Na \textsc{I} doublet (5896 \AA, 5890 \AA) to H$\beta$ for a range of spectral types \citep[using model {\tt Atlas9} data][]{Castelli2004}, overlaid by the EW ratios for all SpUpNIC-observed variables. Model EW ratios increase exponentially towards later spectral types (represented by the dashed curved line), and the measured EW ratio for each of the SpUpNIC variables are shown by a line spanning all spectral types, with the median EW ratio uncertainty indicated by a vertical bar, representing the associated inherent uncertainty in the spectral classification at a given spectral type. Likely spectral classes for each variable are listed in Table~\ref{spectraltype}, ordered by early to late type spectral class (horizontal lines from bottom to top). The seven latest spectral type targets are labeled. }
\label{FIG:SpUpNIC_all}
\end{figure*}

In order to determine the approximate spectral type for each of the targets observed with SpUpNIC, we measured the equivalent width (EW) ratio of the Na \textsc{I} doublet (5890 \AA, 5896 \AA) to H$\beta$ for each target's spectrum, and compared these EW ratios to the EW ratios of several spectral types obtained from  model {\tt Atlas9} data \citep{Castelli2004}. In Table~\ref{spectraltype} we list the EW ratios for each of these OW targets and in Figure~\ref{FIG:SpUpNIC_all} we show where these EW ratios overlap the expected spectral type EW ratio curve. The uncertainties in the individual EW ratios listed in Table 4 give an indication of the uncertainty in the spectral type as derived from Figure 2. In the subsequent figures and tables we only list the approximate spectral type. The SpUpNIC-observed spectra for all targets, ordered by their proposed spectral type, are presented in Figure~\ref{FIG:allspec_1}.

\begin{table*}
\centering
\caption{Equivalent Width ratios of Na \textsc{I} doublet (5896 \AA,
  5890 \AA) to H$\beta$, and approximate spectral type for the SpUpNIC-observed variables}
  \resizebox{\textwidth}{!}{
\begin{tabular}{lrlclrl} 
\hline
SpUpNIC-observed& EW Ratio & $\sim$ Spectral & & SpUpNIC-observed & EW Ratio & $\sim$ Spectral\\
Variable& (Na \textsc{I}/H$\beta$) & Type & &  Variable& (Na \textsc{I}/H$\beta$) & Type \\
\hline 
OW\,J080313.04-320720.2	&	0.058$\pm$0.028	&	F2V	&	&	OW\,J081321.93-313537.7	&	0.112$\pm$0.030	&	F6V/F7V	\\
OW\,J075049.73-270018.3	&	0.061$\pm$0.019	&	F2V	&	&	OW\,J080257.23-292136.0	&	0.113$\pm$0.056	&	F6V/F7V	\\
OW\,J075121.36-270916.4	&	0.061$\pm$0.029	&	F2V	&	&	OW\,J075114.46-270311.2	&	0.117$\pm$0.028	&	F6V/F7V	\\
OW\,J080242.91-314653.4	&	0.067$\pm$0.048	&	F3V/F4V	&	&	OW\,J080517.37-310257.5	&	0.119$\pm$0.019	&	F6V/F7V	\\
OW\,J074216.77-241327.5	&	0.073$\pm$0.026	&	F3V/F4V	&	&	OW\,J080527.94-275144.8	&	0.120$\pm$0.032	&	F6V/F7V	\\
OW\,J080205.41-295804.6	&	0.074$\pm$0.029	&	F3V/F4V	&	&	OW\,J081234.73-312409.8	&	0.120$\pm$0.012	&	F6V/F7V	\\
OW\,J074106.40-293325.5	&	0.074$\pm$0.020	&	F3V/F4V	&	&	OW\,J080146.04-294518.2	&	0.120$\pm$0.047	&	F6V/F7V	\\
OW\,J075837.37-292904.2	&	0.075$\pm$0.038	&	F3V/F4V	&	&	OW\,J075918.25-313721.6	&	0.121$\pm$0.028	&	F6V/F7V	\\
OW\,J074447.99-262155.6	&	0.078$\pm$0.027	&	F3V/F4V	&	&	OW\,J080901.66-291652.2	&	0.123$\pm$0.031	&	F6V/F7V	\\
OW\,J074533.75-263212.5	&	0.079$\pm$0.024	&	F3V/F4V	&	&	OW\,J080205.31-300331.4	&	0.124$\pm$0.017	&	F6V/F7V	\\
OW\,J075359.78-275732.6	&	0.079$\pm$0.027	&	F3V/F4V	&	&	OW\,J075527.61-314825.3	&	0.124$\pm$0.030	&	F6V/F7V	\\
OW\,J080911.11-300209.2	&	0.084$\pm$0.026	&	F5V	&	&	OW\,J074101.18-291540.0	&	0.128$\pm$0.021	&	F8V	\\
OW\,J075952.17-292817.6	&	0.086$\pm$0.022	&	F5V	&	&	OW\,J073649.27-295601.8	&	0.129$\pm$0.042	&	F8V	\\
OW\,J074513.22-261036.0	&	0.087$\pm$0.030	&	F5V	&	&	OW\,J074551.57-242147.8	&	0.130$\pm$0.031	&	F8V	\\
OW\,J080418.75-293740.2	&	0.090$\pm$0.157	&	F5V	&	&	OW\,J075620.67-291605.4	&	0.136$\pm$0.585	&	F8V	\\
OW\,J080916.72-292158.0	&	0.091$\pm$0.044	&	F5V	&	&	OW\,J082257.05-280051.0	&	0.152$\pm$0.024	&	F8V	\\
OW\,J074059.42-242452.4	&	0.092$\pm$0.033	&	F5V	&	&	OW\,J074127.67-243642.9	&	0.152$\pm$0.064	&	F8V	\\
OW\,J080323.23-314852.6	&	0.094$\pm$0.027	&	F5V	&	&	OW\,J080910.70-300744.1	&	0.246$\pm$0.031	&	G1V	\\
OW\,J081018.30-273749.1	&	0.095$\pm$0.025	&	F5V	&	&	OW\,J080626.33-302545.3	&	0.256$\pm$0.032	&	G1V	\\
OW\,J081116.72-313821.1	&	0.096$\pm$0.022	&	F5V	&	&	OW\,J080531.16-274221.5	&	0.276$\pm$0.034	&	G1V	\\
OW\,J075640.83-291107.9	&	0.098$\pm$0.019	&	F5V	&	&	OW\,J080202.79-302356.3	&	0.298$\pm$0.058	&	G2V	\\
OW\,J073909.96-300620.0	&	0.100$\pm$0.023	&	F5V	&	&	OW\,J080154.61-281513.9	&	0.359$\pm$0.046	&	G5V	\\
OW\,J073823.16-303958.0	&	0.101$\pm$0.061	&	F5V	&	&	OW\,J075531.59-315058.2	&	1.253$\pm$0.154	&	late G-type$^{a}$	\\
OW\,J074515.73-260841.7	&	0.108$\pm$0.025	&	F6V/F7V	&	&		&		&		\\
\hline
\multicolumn{7}{r}{$^{a}$determined through inspection of absorption features.}\\
\end{tabular}}
\label{spectraltype}
\end{table*}

From this first-order spectral classification, we determined that the majority of our targets exhibit F-type stellar spectra  with strong Balmer absorption lines and weaker evidence of the Na\,\textsc{I} doublet. There is also evidence for ionized Ca lines at 3968 \AA\ (H) and 3933 \AA\ (K), and (in some cases) the Mg \textsc{I} triplet (5167 \AA, 5172 \AA, 5183 \AA). 
Furthermore, as these F-type stellar systems exhibit short-period variations (P$_{LS} <$ 113 mins), with amplitudes $<$ 0.3 mag, they are therefore most likely to be low-amplitude $\delta$\,Sct stars. 

A number of follow-up targets have been identified as having G-type spectral features, and are therefore unlikely to be $\delta$\,Sct stars Compared to the spectra of targets identified as F-type, the ratio of the Na \textsc{I} doublet to H$\beta$ in each of these G-type spectra is relatively larger, and the Ca lines (H and K) are stronger than all other lines. We stress that our method for determining spectral types using the EW of the Na \textsc{I} doublet is only a first-order approximation. In order to improve the spectral identification of our targets, we could use additional lines for comparison, such as the Ca H and K lines or the Mg triplet lines around 5172 \AA (which is one of the best diagnostics for determining late spectral types, such as G-type stars). In future spectral follow-up observations, we will try to shift the wavelength range bluewards so as to improve the SNR of the Ca H and K lines.

\section{The population of variables}
\label{SEC:varpop}

\begin{figure*}
\centering
\includegraphics[width=\textwidth]{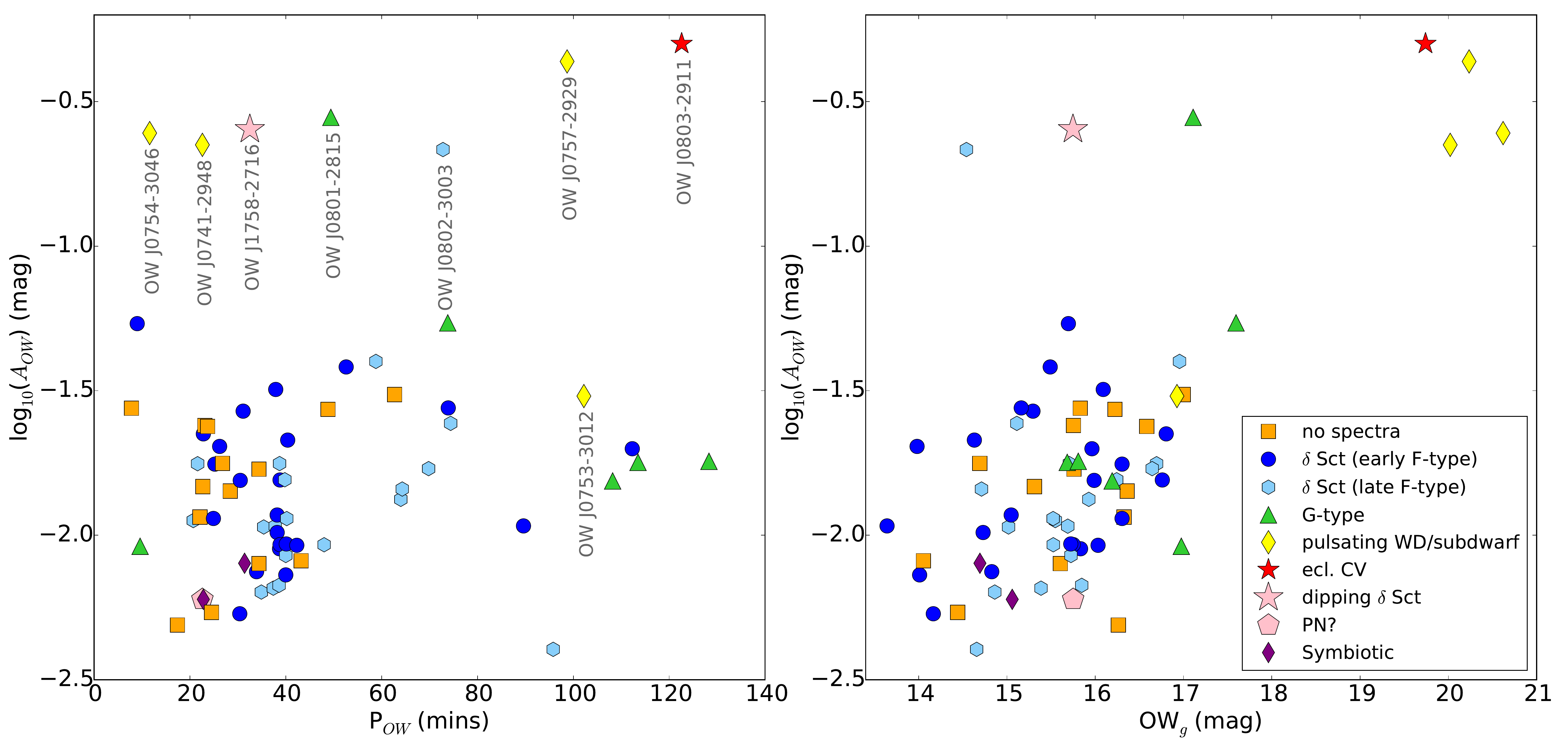}
\caption{All 71 OW targets selected for follow-up observations (with properties listed in Table~\ref{TAB:photproperties} and Table~\ref{TAB:specproperties}) in period-amplitude space, left panel, and magnitude-amplitude space, right panel (according to their OW periods (P$_{OW}$), amplitudes ($A_{OW}$), and Sloan ${g}$ magnitudes (OW$_g$)). For more details of these distributions in OW data, see Paper$\,$II. }
\label{FIG:Pamp}
\end{figure*}

\begin{figure*}
\centering
\subfloat{\label{FIG:colblue}\includegraphics[width=0.69\textwidth]{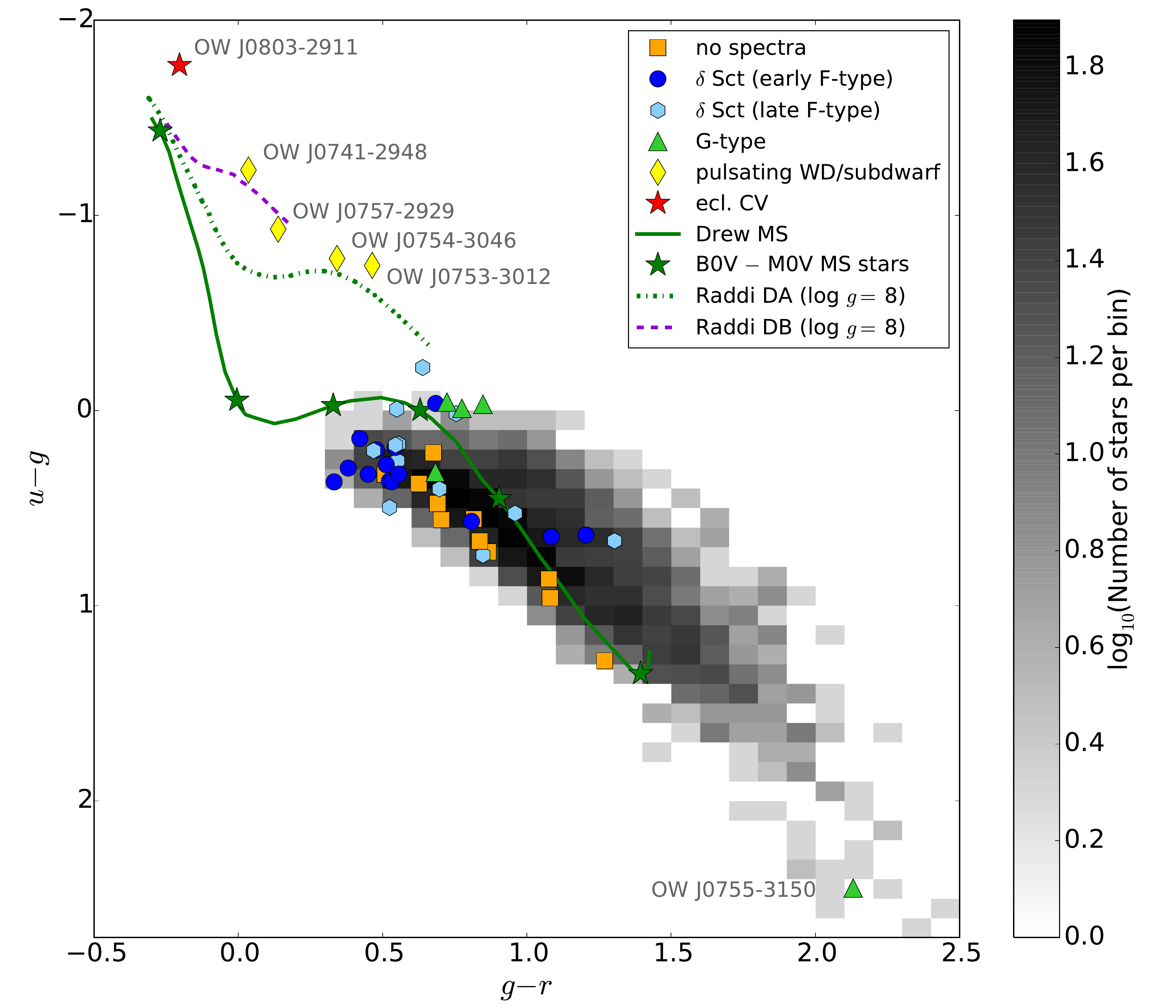}}
\hspace{0.2cm}
\subfloat{\label{FIG:colred}\includegraphics[width=0.69\textwidth]{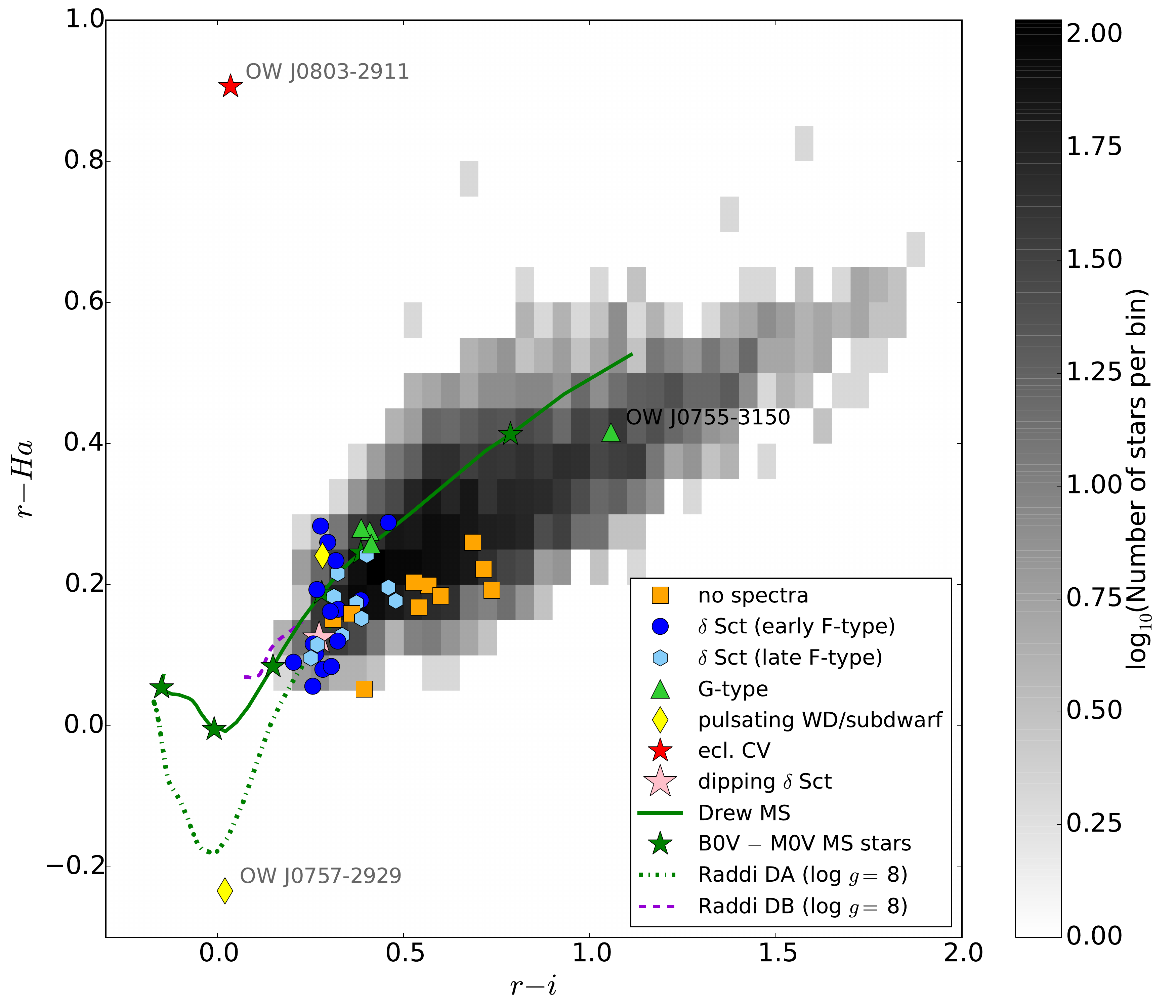}} 
\caption{Distribution of all 71 follow-up targets, for which VPHAS+ colour information was available is shown in $u-g$ vs $g-r$ space, overlapping a 2D histogram of all OW stars with colours indices from ESO P88 - P94 (in bins of 0.1 mag units for the upper panel, and in bins of 0.05 mag units for the lower panel). For the Main Sequence (MS) track, we used the synthetic colours provided by \citet{Drew2014} for R$_v$=3.1 and extinction coefficient A$_0$ = 0, indicated by `Drew MS'. DA and DB tracks are shown, using synthetic colour colour tracks with surface gravities of log $g$ = 8.0 provided by \citet{Raddi2016}. For more details of the colour distributions  in OW data, see Paper$\,$II.}
\label{FIG:colplots}
\end{figure*}

Having established the validity of the OW pipeline-derived amplitudes and periods, and determining the spectral classification for some of our targets, we plot our 71 follow-up targets in period-amplitude and magnitude-amplitude space  in Figure~\ref{FIG:Pamp}, and indicate the proposed nature of the system as determined through spectroscopic identification. Furthermore, by cross-matching our targets with the VST Photometric H$\alpha$ Survey of the Southern Galactic Plane \citep[VPHAS+,][]{Drew2014} in order to obtain multiband colour information, we were able to retrieve $u$-band, $g$-band and $r$-band filter photometry for 48 of our targets, and $r$-band, $i$-band and H$\alpha$-band filter photometry for 43 follow-up targets. In Figure~\ref{FIG:colplots}, we show  the position of these objects in two colour-colour planes and, as in Figure~\ref{FIG:Pamp}, we show the spectral classification of the targets where possible, and label those systems with only photometric follow-up as `no spectra'.

As shown in the left panel of Figure~\ref{FIG:Pamp}, most of our stars appear to clump together at periods between 20 and 45 minutes with amplitudes $A_{OW} <$ 0.04 mag ($\log_{10}(A_{OW}) \leq$ --1.4). The majority of these objects have been spectroscopically identified as early/late F-type stars and are thus likely low-amplitude $\delta$\,Sct stars.  Furthermore, their positions in colour-colour space are consistent with the known spectral types of $\delta$\,Sct stars \citep{Chang2013}. Although the remaining stars within this region have only been observed photometrically, their location within this plane implies that they are likely to be $\delta$\,Sct stars. However, their precise nature remains uncertain as the two symbiotic binaries and the central star of the planetary nebula candidate also lie within this region. 

The second largest clump of stars is visible with periods between 45 mins and 80 mins and with amplitudes $A_{OW} <$ 0.04 mag. With the exception of one star (OW$\,$J080531.16--274221.5), these targets are identified as low-amplitude F-type $\delta$\,Sct stars. By plotting all the $\delta$\,Sct stars in colour-colour space (as shown in Figure~\ref{FIG:colplots}), we can see that these classifications largely agree with the expected (reddened) stellar colours (see also Paper\,II). OW$\,$J080531.16--274221.5, itself, exhibits the spectral features of a G1V-type spectrum. Most of the other targets identified with G-type spectra have variations with relatively longer periods (in the region where P$_{OW} >$ 100 mins and $A_{OW} <$ 0.04 mag).  

Four of the stars, namely OW$\,$J181847.29--251231.9, OW$\,$J073823.16--303958.0, OW$\,$J075531.59--315058.2, and OW$\,$J073919.72--300922.7, exhibit modulations with periods under 20 minutes, and amplitudes $A_{OW}$ ranging from 0.005 mag to 0.06 mag (--2.3 $\leq \log_{10}(A_{OW}) \leq$ --1.2 ). Apart from OW$\,$J181847.29--251231.9, the period of these systems still needs to be validated using follow-up photometric observations. If these short periods can be supported by additional data, then we would be populating the short-period end of the period distribution of $\delta$\,Sct stars \citep[][find a period minimum of $\sim$18 mins in Kepler observations of $\delta$\,Sct stars]{Uytterhoeven2011}. Furthermore, their positioning in period-amplitude space overlaps with the location of objects such as rapidly oscillating Ap stars (roAp), and pulsating hydrogen-atmosphere (DA) white dwarfs (ZZ Ceti). However, OW$\,$J073823.16--303958.0 exhibits an early F-type spectrum, and OW$\,$J075531.59--315058.2 is a very red object (as shown in Figure~\ref{FIG:colblue}), with  a late G5V-type spectrum so both are not DA white dwarfs.  

\begin{figure*}
\centering
\includegraphics[width=\linewidth]{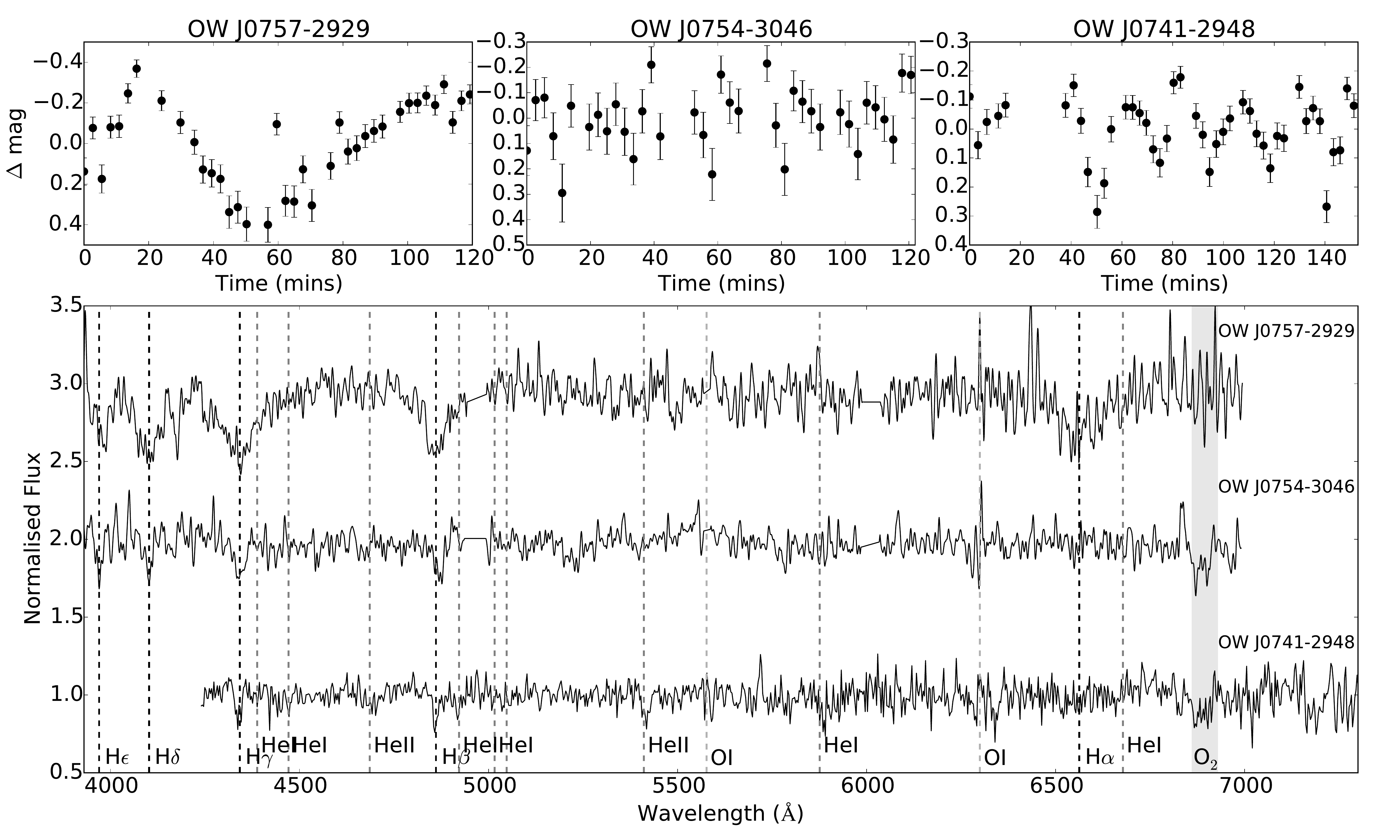}
\caption{$g-$band OW light curves (upper panels), and continuum-normalised SALT-observed spectra (lower panel) of the three pulsating white dwarf, subdwarf, or binary star candidates:  OW$\,$J075719.94--292955.5, OW$\,$J075436.40--304655.0, and OW$\,$J074106.07--294811.0. See Section~\ref{SEC:UCBsearch} for details. Dashed lines indicate important spectral lines (as labeled), and shaded areas cover atmospheric telluric bands.}
\label{FIG:pulsatingWD}
\end{figure*}

Lastly, seven of the targets have relatively large amplitudes, greater than 0.1 mag ($\log_{10}(A_{OW}) \geq$ --1.0). In addition to a possible high-amplitude $\delta$\,Sct star (OW$\,$J080205.31--300331.4), and a G5V stellar object (OW$\,$J080154.61--281513.9), this subset includes three possible pulsating white dwarfs/subdwarfs (OW$\,$J075436.40--304655.0, and OW$\,$J074106.07--294811.0, and OW$\,$J075719.94--292955.5), an eclipsing CV candidate (OW$\,$J080311.83--291144.6), and a high-amplitude $\delta$\,Sct star candidate (OW\,J175848.21--271653.7), which also appears to exhibit small dipping features in its SHOC follow-up light curve. The latter five targets are discussed further in Section~\ref{SEC:UCBsearch} and Section~\ref{SEC:Eclipse}. 

\section{Searching for Ultra Compact Binaries}
\label{SEC:UCBsearch}
\begin{figure*}
\centering
\includegraphics[width=\linewidth]{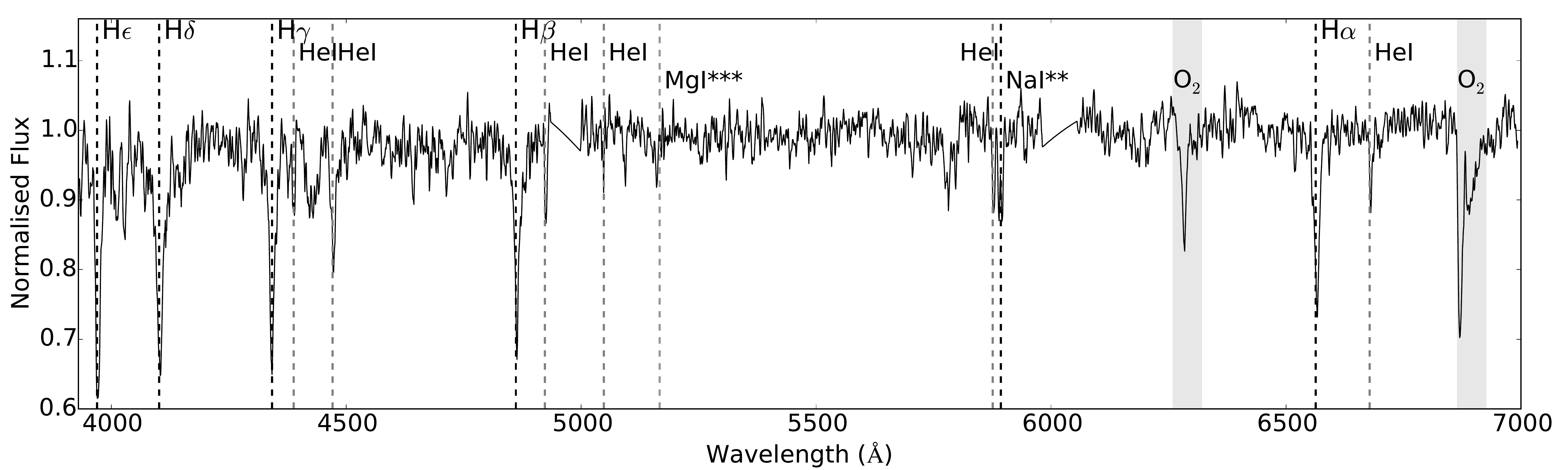}
\caption{Normalised SALT-observed spectrum of early B-type star or binary system candidate OW\,J075305.78--301208.1. Dashed lines indicate important spectral lines (as labeled), and shaded areas cover atmospheric telluric bands.}
\label{FIG:OW0753}
\end{figure*}

Four of the bluest targets, namely OW$\,$J074106.07--294811.0, OW$\,$J075719.94--292955.5, OW$\,$J075436.40--304655.0, and OW\,J075305.78--301208.1 (hereafter OW$\,$J0741--2948, OW$\,$J0757--2929, OW$\,$J0754--3046 and OW\,J0753--3012 resp.) were observed by SALT as our most likely UCB candidates, exhibiting colours most consistent with known AM\,CVn systems. However, only one of these systems, OW$\,$J0741--2948, has been found to be a UCB. Additional photometry and spectroscopy is needed to further classify these four systems. In Figure~\ref{FIG:pulsatingWD}, we compare the SALT-observed spectra of two other UCB candidates, namely OW$\,$J0757--2929 and OW$\,$J0754--3046, to the spectrum of OW$\,$J0741--2948, and in Figure~\ref{FIG:OW0753}, we show the spectrum of OW\,J0753--3012.

\subsection{The UCB OW$\,$J0741--2948}

As discussed in Paper$\,$I, the SALT-observed spectrum of OW$\,$J0741--2948 revealed weak narrow Balmer absorption lines and little evidence of helium. Further photometric and spectroscopic follow-up of this target (shown in Figure~\ref{FIG:pulsatingWD}) reveals it is in a binary system with a 44-minute orbital period, consisting of a hot subdwarf O star and a compact companion. Additional follow-up data of this target has now been obtained and a future paper, Kupfer et al. (in prep),  will be dedicated to exploring the properties of this unique target.

\subsection{OW$\,$J0757--2929}

OW$\,$J0757--2929 exhibits the broad Balmer absorption lines characteristic of a DA white dwarf pulsator, with little evidence of helium or other metals (although the low SNR in all spectra in Figure~\ref{FIG:pulsatingWD} make line identification other than the hydrogen lines uncertain). Furthermore, the blue colours of this target are indicative of a massive hot DA white dwarf ( $u-g$ $=$ --0.93, $g-r$ $=$ 0.14 for OW$\,$J0757--2929).  However, in the OW light curve of OW$\,$J0757--2929, the amplitude ($A_{OW} =$ 0.44 mag) is larger than what is expected in a pulsating or rotating white dwarf. Also, there is no evidence for the short-periods typically associated with DA white dwarf pulsations. Instead it exhibits a modulation on a period longer than the observation duration ($>$ 2 hours). It is possible that OW$\,$J0757--2929 is a binary system consisting of a hot white dwarf and a M-dwarf binary. In this case, the variations could be due to the reflection effect from the cool M-dwarf companion on the orbital period. Another possibility is that OW$\,$J0757--2929 is a cataclysmic variable that was in a low- or no-accretion state when the spectrum was observed. 

\subsection{OW$\,$J0754--3046}
In the case of OW$\,$J0754--3046, there is no evidence of H$\alpha$ absorption which could either be due to intrinsic weakness of the line, or (partial) infill by emission (similar to what is seen in OW$\,$J0741--2948). Furthermore, the absorption lines are narrower than lines typically associated with white dwarfs. The spectrum more closely resembles that of a hot pulsating subdwarf or a distant B star, although the large photometric amplitude ($A_{OW} =$ 0.25 mag) suggests the variations do not originate from pulsations. As in OW$\,$J0757--2929, OW$\,$J0754--3046 exhibits very blue colours ($u - g$ $=$--0.78, $g-r$ $=$ 0.34). A period of 11.6 min was identified in the light curve of OW J075436-30 using the Lomb Scargle test. The OW pipeline uses difference imaging to obtain light curves for each star in our images (see Paper I for details). However, as part of our process to verify candidate variables identified using the pipeline (see Paper II for further
details), additional differential aperture photometry did not confirm
the 11.6 min period, but some evidence was found for a longer period
in this source. 

\subsection{OW\,J0753--3012}
The target OW\,J0753--3012 was selected for SALT follow-up based on its very blue colour ($u - g$ $=$--0.74, $g-r$ $=$ 0.46). The SALT spectrum of this target, shown in Figure~\ref{FIG:OW0753}, reveals absorption lines of H I, He I, the Na I doublet, and some weak evidence for the Mg I triplet. Since He I $\lambda$5876 and the Na I doublet are of equal strength, the He I absorption lines are relatively strong, and there is no clear evidence for He II absorption, this implies the source has the spectral type of an early B-type star. As OW\,J0753--3012 also exhibits a long period variation in its OW light curve ($P_{OW} \sim$ 102 mins), it is possibly a slow or hybrid subdwarf B variable. Detailed photometric observations of this target are now required to confirm the modulation period detected by OW, and to search for the shorter-period variations typically associated with hybrid sdB stars.  

\begin{figure*}
\centering
\includegraphics[width=\linewidth]{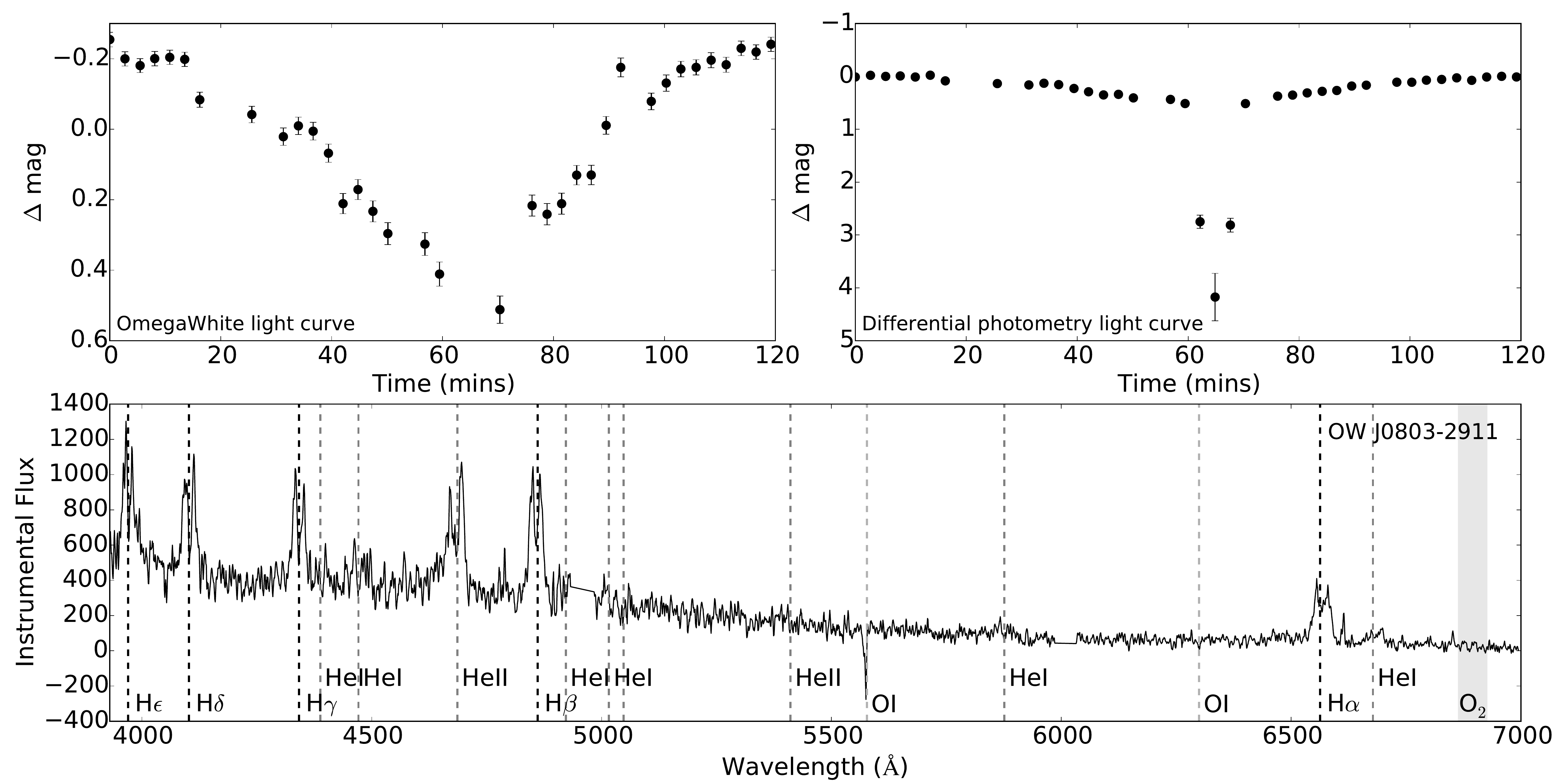}
\caption{$g-$band OW light curve (upper left panel), differential photometry (upper right panel), and SALT-observed spectrum (lower panel) of eclipsing CV candidate OW\,J080311.83--291144.6. Dashed lines indicate important spectral lines (as labeled), and shaded areas cover atmospheric telluric bands.}
\label{FIG:OW0803}
\end{figure*}

\section{The Eclipsing System OW\,J080311.83--291144.6}
\label{SEC:Eclipse}

The target OW\,J080311.83--291144.6 (hereafter OW\,J0803--2911), was selected for SALT spectroscopic follow-up, as it exhibited very blue colours ($u-g$ $=$ --1.77, $g-r$ $=$ --0.20). According to the colour-colour plots in Figure~\ref{FIG:colplots}, it is the bluest object of all targets with VPHAS+ colours, and additional differential photometry of this target reveal deep eclipses of $<$10.7 minutes duration in its light curve, with an orbital period that is greater than 60 min (see Figure~\ref{FIG:OW0803}, and Paper$\,$II for details). The SALT spectrum for OW\,J0803--2911 exhibits strong double-peaked Balmer emission lines, characteristic of an accretion disc in a cataclysmic variable (CV), and a strong double-peaked He II $\lambda$4686 line. Furthermore, these lines are split deeply, indicating that the system is at a high inclination. Thus, with an orbital period that is greater than the observation duration ($>$ 2 hrs), it can be classified as a cataclysmic variable binary star system.

\section{$\delta$\,Sct-type pulsators}

$\delta$ Sct-type stellar pulsators, and their cousins the SX Phe and
$\gamma$ Dor stars, are amongst the most common of variable stars \citep[e.g.][]{Breger2000}. In Paper II, we found nearly 400
$\delta$ Sct-type candidates based on their photometric variability
properties and $ugr$ colours. In this paper we present optical spectra
of a sub-sample of these $\delta$ Sct-type candidates. 

\subsection{OW$\,$J172235.29-321403.6}
\begin{figure*}
\centering
\includegraphics[width=\linewidth]{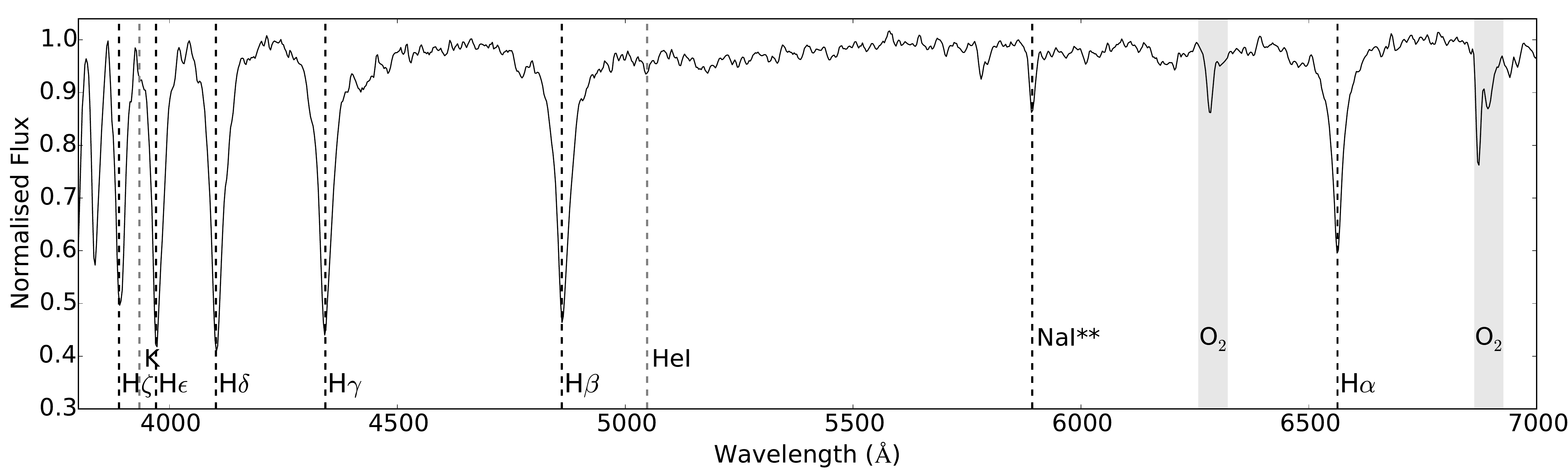}
\caption{Continuum-normalised spectra of the SpUpNIC-observed target OW$\,$J172235.29-321403.6. The target is identified as a planetary nebula candidate in the SIMBAD database (PN\,Th\,3--8), but no emission lines can be seen. Dashed lines indicate important spectral lines (as labeled), and shaded areas cover atmospheric telluric bands.}
\label{FIG:PN}
\end{figure*}

The target OW$\,$J172235.29-321403.6 is identified in SIMBAD as a planetary nebulae candidate (PN Th 3-8). Although the source is classified as a planetary nebula in the SIMBAD database, \citet{Acker1987} find no evidence for emission lines. Similarly, we find no emission line evidence, as shown in the SpUpNIC spectrum of the PN\,Th\,3--8 in Figure~\ref{FIG:PN}. Instead, it exhibits strong Balmer absorption lines and a weak Ca K line, comparable to the spectra of our early F-type $\delta$ Sct stars. We concur with \citet{Acker1987} that this object is misclassified and propose that the classification of PN is removed from SIMBAD. 

\subsection{The Unique $\delta$\,Sct-type candidate OW\,J175848.21--271653.7}
\label{J1758}

\begin{figure*}
\centering
\includegraphics[width=\linewidth]{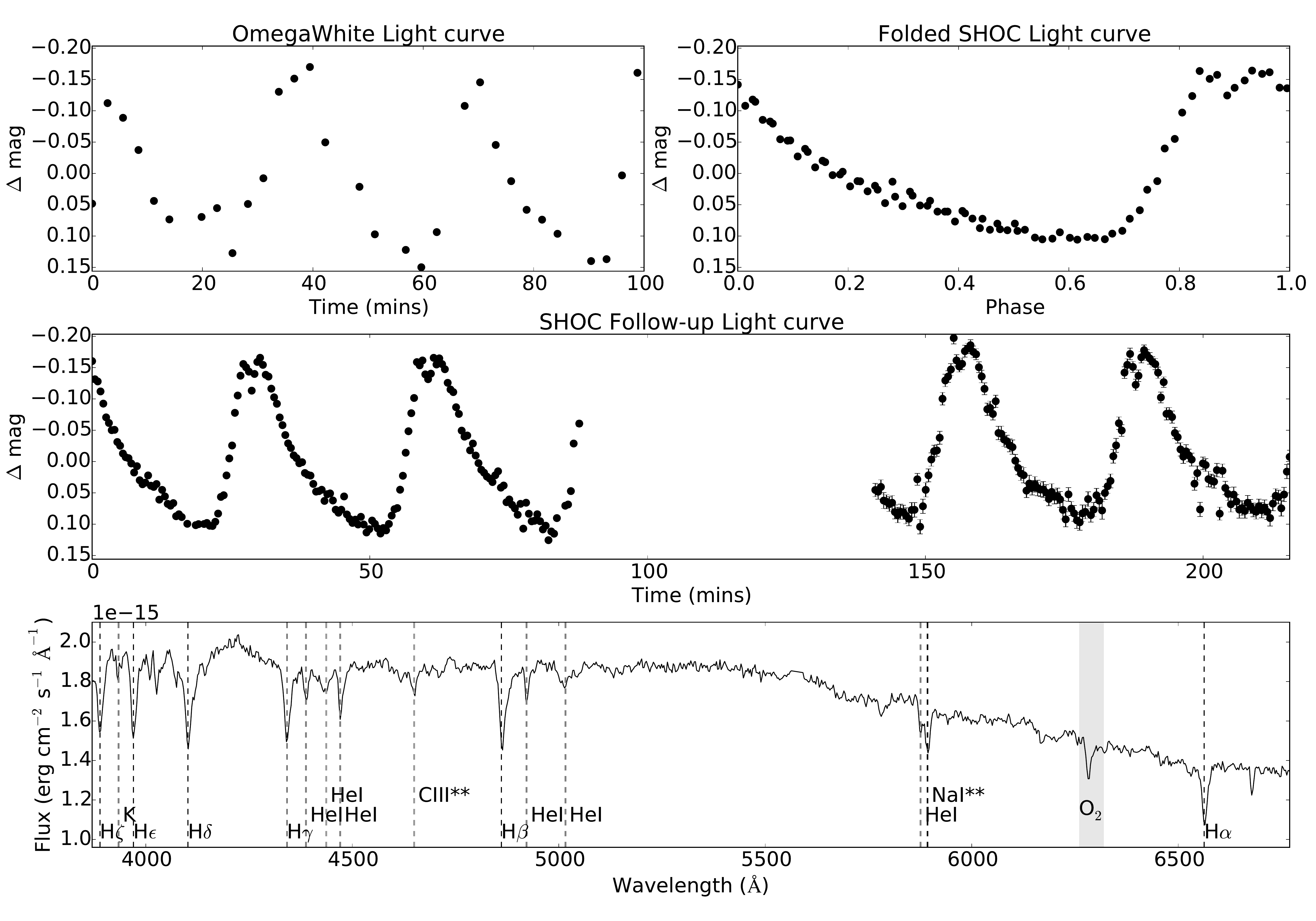}
\caption{$g-$band OW light curve (upper left panel), and the associated SHOC light curve (no filter), the first two cycles folded on $P_{SHOC}$ (upper right panel) and all the data unfolded (middle panel) of $\delta$\,Sct-type candidate OW\,J175848.21--271653.7. SHOC data has been binned in either 1-minute (folded data), or 30-s (unfolded data) intervals. The SpUpNIC-observed spectrum is shown in the lower panel, with important spectral lines indicated, and shaded areas covering either atmospheric telluric bands or the G-band (CH).}
\label{Fig_31a_70_204687}
\end{figure*}

The photometry of OW\,J175848.21--271653.7 (hereafter OW\,J1758--2716) shows characteristics typical of a $\delta$\,Sct-type pulsator but with an unusual dip at the peak of the pulsation cycle. This relatively bright target (OW$_{g} \approx$ 15.75 mag) was originally selected for follow-up observations based on its short-period oscillations (P$_{OW}$ $\approx$ 32.5 mins), and its low false alarm probability ($\log_{10}$(FAP) $\approx$ --3.49). Furthermore, the OW light curve of this target exhibits clear modulations with an amplitude of 0.254 mag. The asymmetric light curve shape, typical of $\delta$\,Sct-type light curves, is clearly observed in the follow-up SHOC light curves. However, the SHOC light curves also reveal a dip in magnitude during peak brightness in each cycle, as shown in Figure~\ref{Fig_31a_70_204687}.\\
The dip, which is 22\% of the peak-to-peak amplitude with a duration of $\sim$3 minutes, has never been seen before in $\delta$\,Sct stars (to the best of our knowledge). However, it is very reminiscent of the effect predicted and seen in Cepheid stars, related to an opacity/ionization dip in so-called `Bump Cepheids' (\citealt{Adams1978}; \citealt{Bono2000}; \citealt{Mihalas2003}). This occurs when the ionization shock front driving the pulsation is exposed in the photosphere of the star, due to a resonance between the fundamental mode and the second radial overtone having a period ratio of exactly 2. As part of the radiated energy is being absorbed in the ionization zone it causes a decrease of the luminosity of the star. Simulations of Cepheids show that the depth and phase at which this occurs is a sensitive parameter for the mass and evolutionary state of the system (e.g. the 7.4 M$_\odot$ Cepheid model in Fig. 3 in \citet{Adams1978} is almost a carbon-copy of our light curve for OW\,J1758--2716). Such modelling for $\delta$ Sct stars is currently lacking, but would allow a determination of their internal structure.
However, the $\delta$ Sct star classification has now been challenged by spectroscopic observations. The SpUpNIC spectra of OW\,J1758--2716, shown in Figure~\ref{Fig_31a_70_204687}, reveals a combination of strong H I and He I absorption lines (as well as some evidence for K and the Na I and C III doublets). This spectra appears to indicate that OW\,J1758--2716 is a B-type star, rather then the A- or F-type star that is associated with $\delta$ Sct stars. Detailed observations are now required to determine the precise nature of this intriguing object.

.
\section{The Symbiotic Binaries SS73\,122 and Hen\,2--357}
\label{SIMBAD}

\begin{figure*}
\centering
\includegraphics[width=\linewidth]{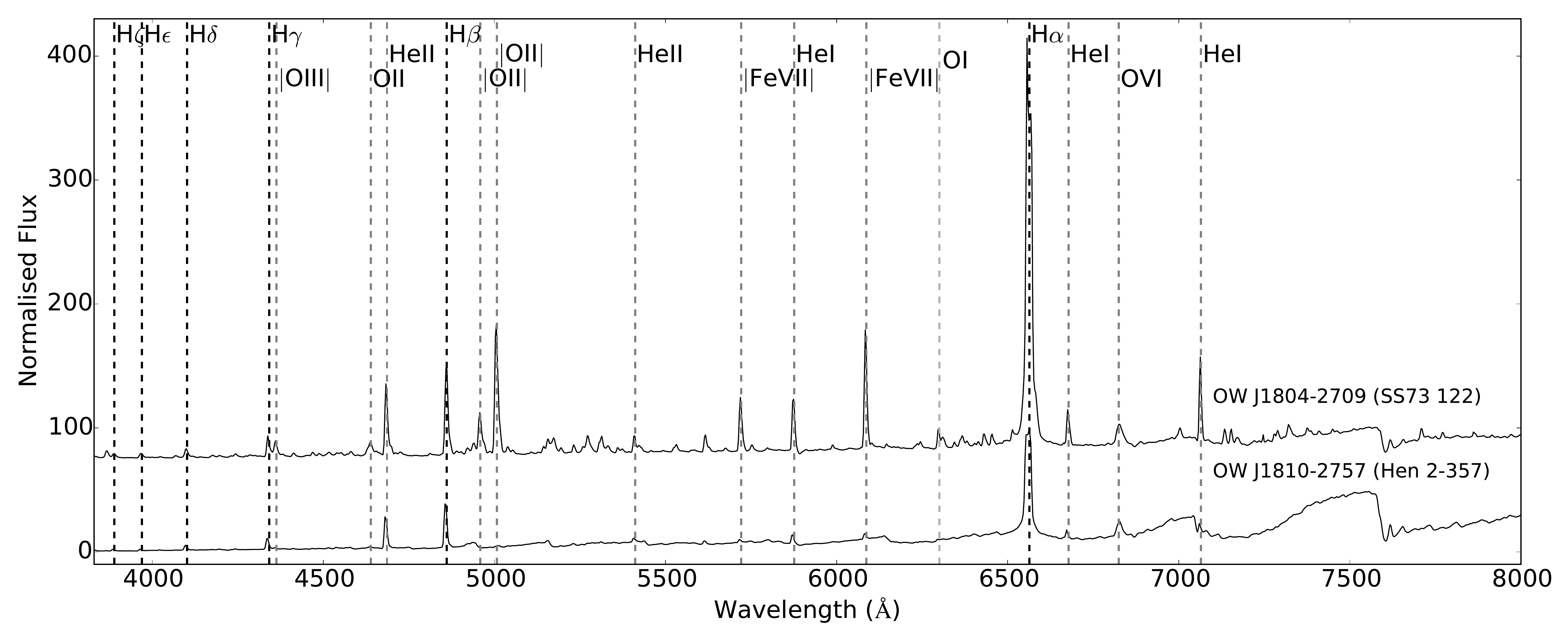}
\caption{Continuum-normalised spectra of the SpUpNIC-observed targets OW$\,$J180441.23-270912.4 and OW$\,$J181043.86-275750.0. The targets are each identified as symbiotic binary stars in the SIMBAD database (SS73\,122 and Hen\,2--357 resp.). An offset of 75 units has been applied to the spectrum of SS73\,122 for visualisation purposes. Dashed lines indicate important spectral lines (as labeled).}
\label{FIG:Sym}
\end{figure*}

As discussed in Paper$\,$II, a number of short-period variables from the first 4 years of OW observations have been identified in the SIMBAD database. Two of our spectrographic targets, OW$\,$J180441.23-270912.4 and OW$\,$J181043.86-275750.0 are classified as symbiotic stars in SIMBAD (SS73\,122 and Hen\,2--357 resp.)

The OW light curves of the symbiotic binaries SS73\,122 and Hen\,2--357 show short-periodic variations of 22.8 mins and 31.4 mins respectively. The 22.8 min variation seen in Hen 2-357 is discussed more thoroughly in Paper II, which presents evidence that Hen 2-357 could belong to the rare class of magnetic symbiotic stars. The SpUpNIC-observed spectra for the two sources are shown in Figure~\ref{FIG:Sym}. Both spectra exhibit clear evidence for a late-type giant (e.g. TiO, Ca II absorption bands), and strong emission lines of H I, He I, $|$O II$|$, and  $|$Fe VII$|$ (with weaker evidence for $|$O III$|$ and He II). Furthermore, the Raman-scattered O\,VI emission feature (at $\lambda$ 6823) can be identified in both spectra. According to \citet{Belczynski2000}, this is sufficient to classify them as symbiotic stars. However, in both sources it is apparent that there are additional emission features that can possibly be attributed to a planetary nebulae shell in both cases. This can also be seen in the extended emission observed in the 2D spectra of both targets.

\section{Summary and Conclusions}
\label{SEC:discussion}

In this paper, we present the photometric and spectroscopic follow-up observations of 71 short-period targets, selected from the OW Survey. We have observed 27 of the follow-up targets using SHOC on the SAAO 1.0-m and 1.9-m telescopes, and 57 of the targets have spectroscopic observations, using either SpUpNIC on the SAAO 1.9-m, or the RSS on SALT.

From the analysis presented in Section~\ref{SEC:Photcomp} we conclude that the OW pipeline is delivering periods and amplitudes in concordance to the expectations from simulations presented in Paper$\,$I. A small number of objects in our sample do not show the expected variability behaviour in their SHOC light curves. This can be due to factors such as poor photometry, limited SNR values, long time-scale changes, phasing phenomena, or due to the relatively short duration of the OW light curves. If we exclude the four targets that have the greatest difference between their respective OW and SHOC periods (OW J073919.72--300922.7, OW J080205.41--295804.6, OW$\,$J174308.60-255102.6, and OW J080257.23--292136.0), the OW periods are accurate to within 5\%, with an RMS-scatter of 10\%. This agrees with the accuracies derived in Paper$\,$I for simulated simple sinusoidal light curves for periods at least below 80 minutes. 

The analysis from Section~\ref{SEC:SpecResults} and Section~\ref{SEC:varpop} indicates that low-amplitude $\delta$\,Sct stars constitute the majority of our follow-up targets and that, most importantly, these stand out in period, amplitude and colour planes. In future papers, we will use this sample to better understand the period and amplitude distributions of $\delta$\,Sct stars, as discussed in Paper$\,$II. We can also use these groupings in period/amplitude/magnitude space to bias target selection for future follow-up observations, based on their positions within these planes. We will next target the regions where objects exhibit periods less than 20 mins (possible UCB or pulsating white dwarf candidates), and objects with amplitudes greater than 0.2 magnitudes (where the UCB candidates and eclipsing CV candidate from Section~\ref{SEC:UCBsearch} and Section~\ref{SEC:Eclipse} are located).  

It is clear that the OW Survey is successfully able to provide numerous interesting and unusual objects. The discovery of OW\,J1758--2716 shows that a survey such as OW, coupled to appropriate follow-up photometry, is capable of picking out rare short-period variables. The detection of an eclipsing CV system (OW\,J0803--2911) proves that it is also possible to detect longer-period variables, as well as eclipsing systems, in the  OW survey.  Techniques for detecting transient events such as eclipsing features in OW sources are currently being investigated and will be presented in future OW papers. In addition to these interesting targets, three potentially very rare pulsating subdwarf/white dwarf sources have been detected and now require future dedicated follow-up observations. Similarly, the two SIMBAD symbiotic stars would greatly benefit from high cadence photometric observations in order to confirm their light curve modulation period, and whether these variations are indeed the result of pulsations or rotations. So far, at least one of these symbiotics, Hen\,2--357, exhibit a verified short-period modulation, implying it could be only the second known magnetic symbiotic star. 

As the OW Survey is ongoing, we shall continue to investigate these targets as well any new objects, while our search for the elusive short-period AM CVn binary star systems continues.   

\section*{Acknowledgements}
 The authors gratefully acknowledge funding from the Erasmus Mundus
Programme SAPIENT, the National Research Foundation of South Africa
(NRF), the Nederlandse Organisatie voor Wetenschappelijk Onderzoek
(the Dutch Organisation for Science Research), Radboud University, 
the University of Cape Town. Armagh Observatory is core funded by the Northern Ireland Executive. The ESO observations used in this paper are based on observations made with ESO Telescopes at the La Silla Paranal Observatory under programme IDs: 088.D-4010(B), 090.D-0703(A), 090.D-0703(B), 091.D-0716(A), 091.D-0716(B), 092.D-0853(B), 093.D-0937(A), 093.D-0753(A),  094.D-0502(A), and 094.D-0502(B) as part of the Dutch GTO time on OmegaCAM, and  177.D-3023 (VPHAS+). This paper uses observations made at the South African Astronomical Observatory (SAAO). Some of the observations reported in this paper were obtained with the Southern African Large Telescope(SALT) under programs 2014-2-SCI-030 and 2015-2-SCI-035 (PI: Sally Macfarlane). This research is supported by the NWO/NRF Bilateral agreement supporting astronomical research. This research has made use of the SIMBAD database,
operated at CDS, Strasbourg, France. We thank Sterl Phinney for the  stimulating discussions on the Bump Cepheid phenomena. Furthermore, we thank Thomas Kupfer and the anonymous referee for the useful comments which have helped to improve the paper. Our co-author, Darragh O'Donoghue, died on 25 June 2015 leaving a huge legacy at the SAAO including providing the optical design for the SpUpNIC spectrograph. 


\bibliographystyle{mnras}
\bibliography{references} 


\appendix

\section{Target Light curves}

\begin{figure*}
\centering
\includegraphics[width=\linewidth]{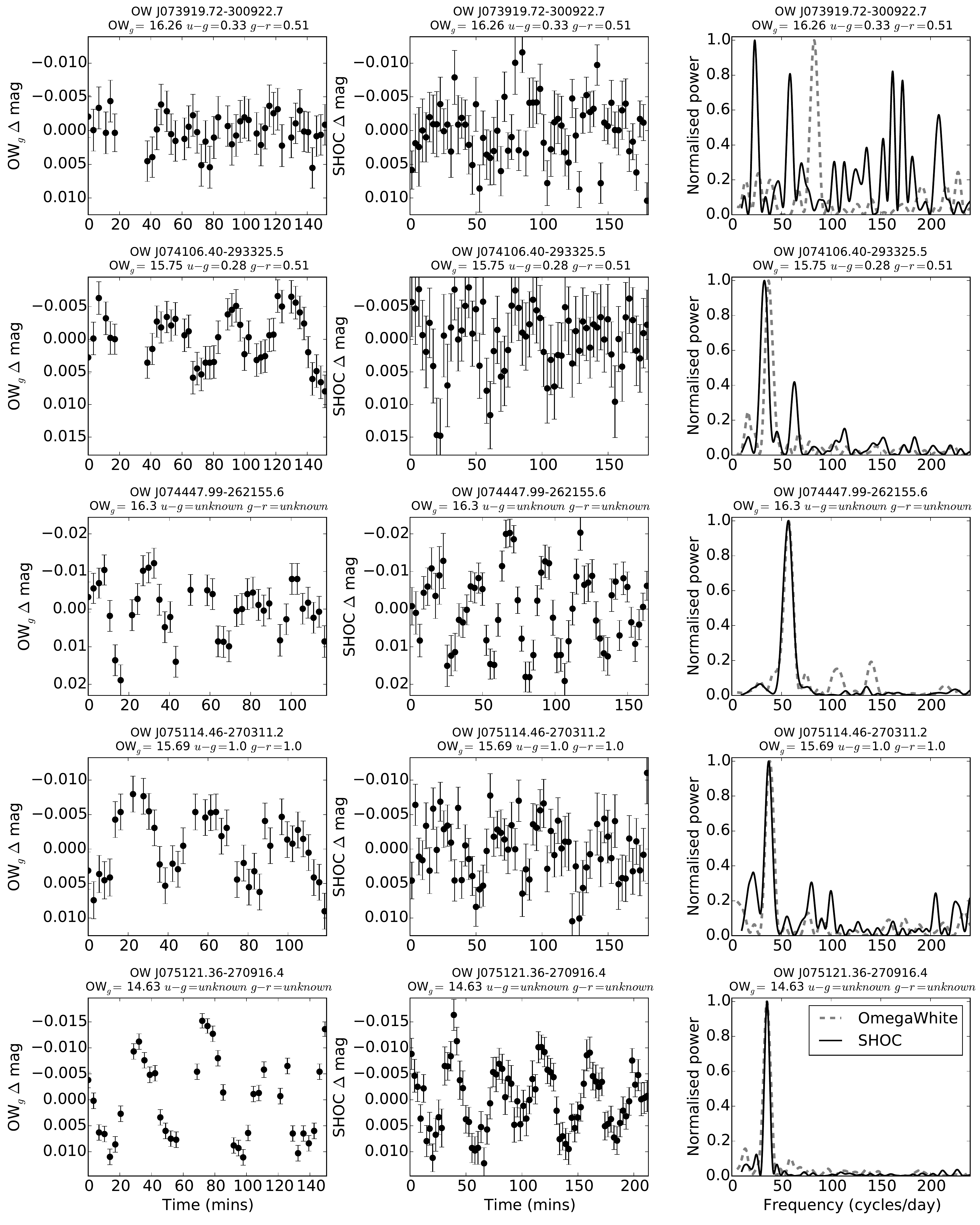}
\caption{OW light curves (left panels), follow-up SHOC light curves (middle panels) and discrete Fourier Transform power spectra (`DFT', right panels) of the selected targets (one target per row, ordered by RA). OW light curves are in g-band and are processed using difference imaging in the OW pipeline. SHOC light curves are observed with no filter, and are binned to match the OW cadence. In each SHOC and OW DFT, the power has been normalised to the peak SHOC or peak OW power resp, for ease of period comparison. See Table~\ref{TAB:validity} for the peak amplitude and associated period for each target.}
\label{FIG_lcft}
\end{figure*}

\begin{figure*}
\centering
\includegraphics[width=\linewidth]{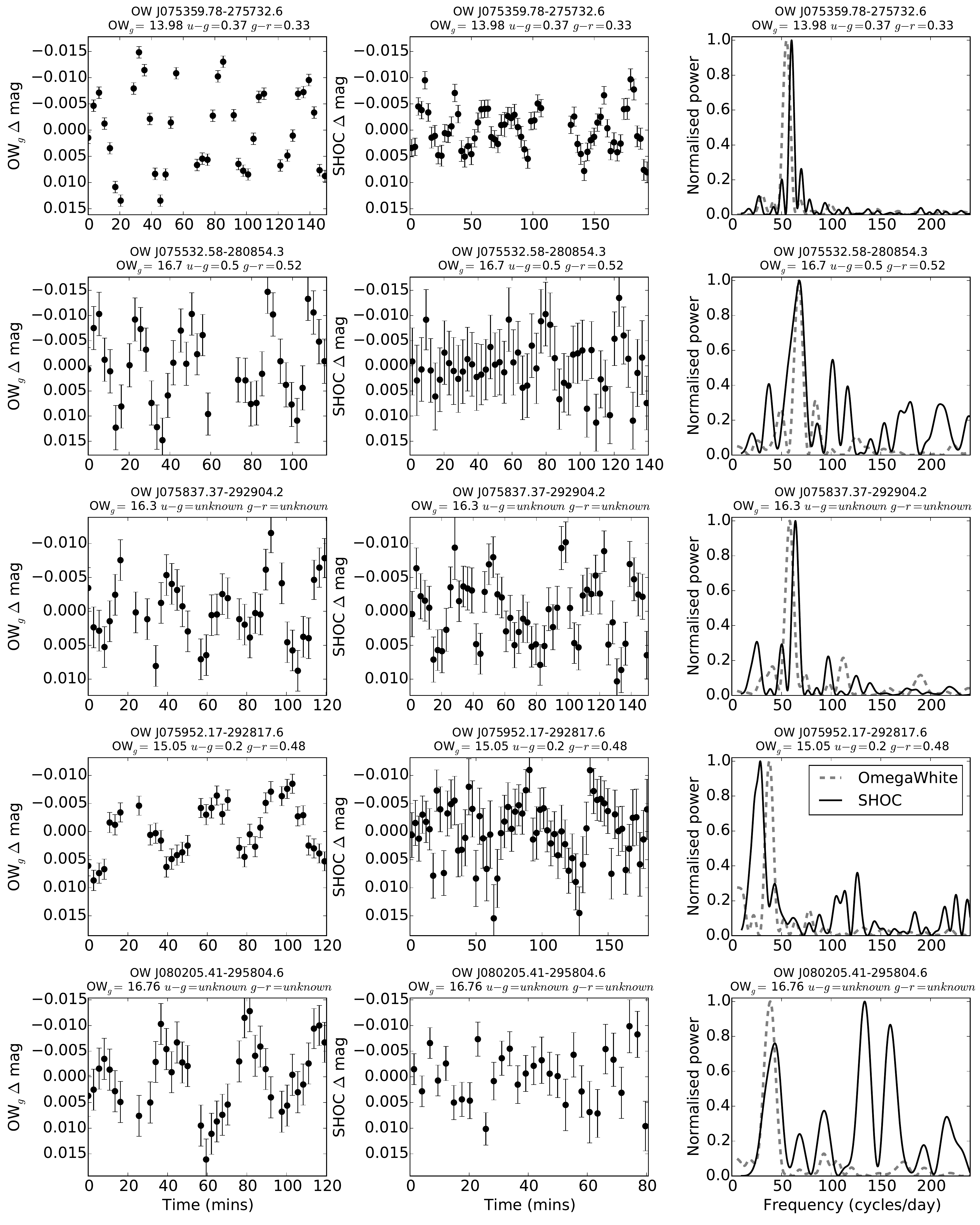}
\contcaption{OW light curves (left panels), follow-up SHOC light curves (middle panels) and discrete Fourier Transform power spectra (`DFT', right panels) of the selected targets (one target per row, ordered by RA). OW light curves are in g-band and are processed using difference imaging in the OW pipeline. SHOC light curves are observed with no filter, and are binned to match the OW cadence. In each SHOC and OW DFT, the power has been normalised to the peak SHOC or peak OW power resp, for ease of period comparison. See Table~\ref{TAB:validity} for the peak amplitude and associated period for each target.}
\end{figure*}

\begin{figure*}
\centering
\includegraphics[width=\linewidth]{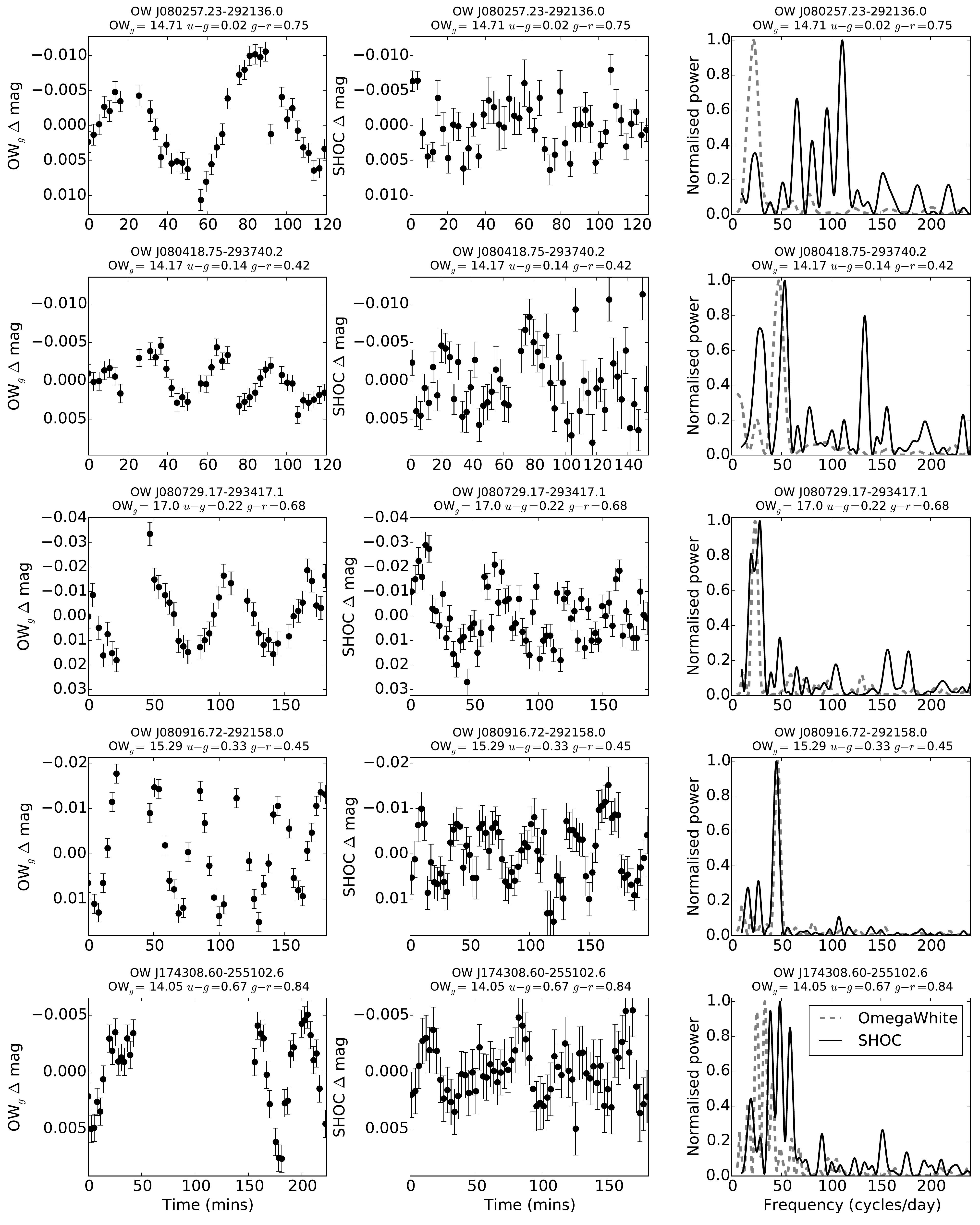}
\contcaption{OW light curves (left panels), follow-up SHOC light curves (middle panels) and discrete Fourier Transform power spectra (`DFT', right panels) of the selected targets (one target per row, ordered by RA). OW light curves are in g-band and are processed using difference imaging in the OW pipeline. SHOC light curves are observed with no filter, and are binned to match the OW cadence. In each SHOC and OW DFT, the power has been normalised to the peak SHOC or peak OW power resp, for ease of period comparison. See Table~\ref{TAB:validity} for the peak amplitude and associated period for each target.}
\end{figure*}

\begin{figure*}
\centering
\includegraphics[width=\linewidth]{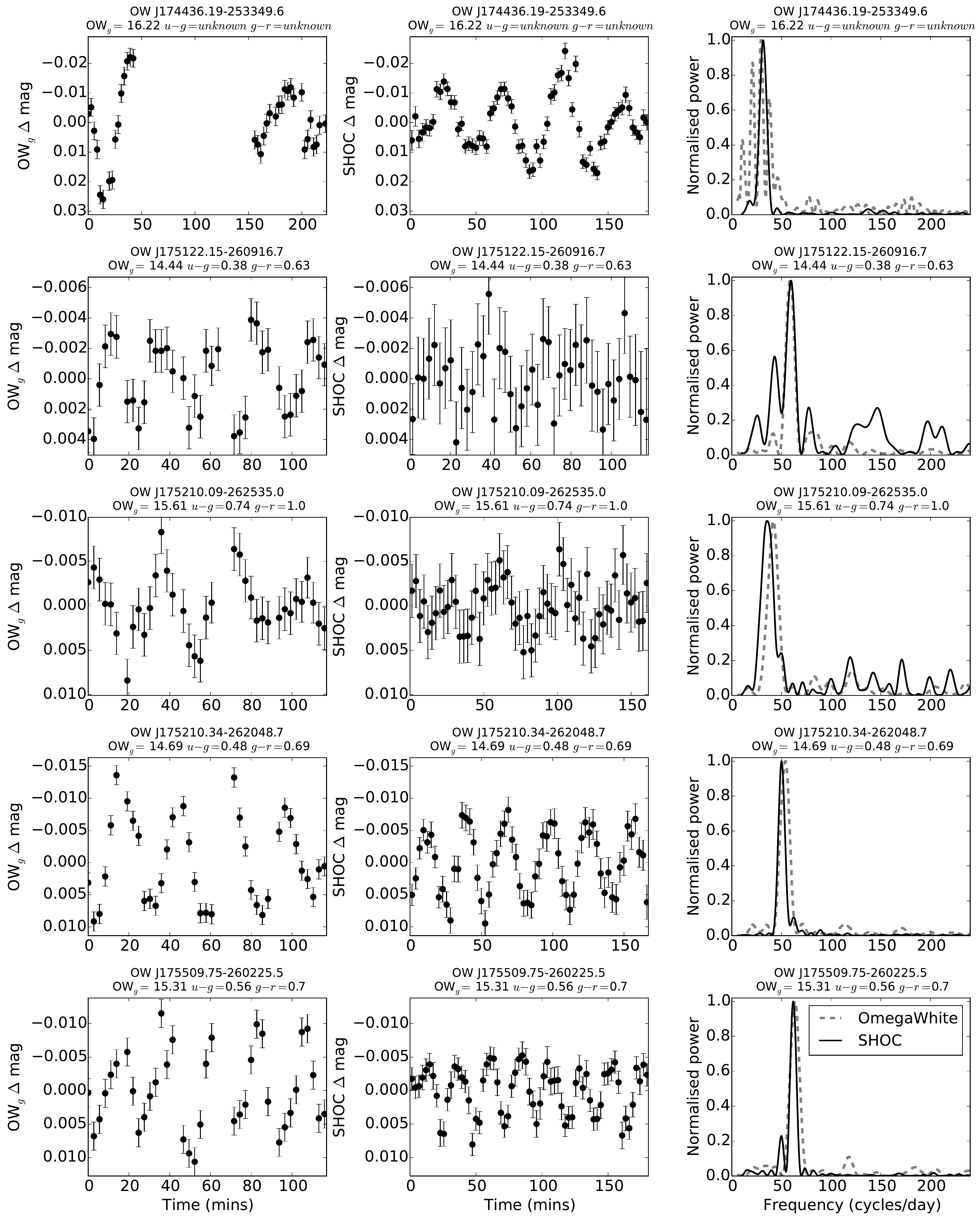}
\contcaption{OW light curves (left panels), follow-up SHOC light curves (middle panels) and discrete Fourier Transform power spectra (`DFT', right panels) of the selected targets (one target per row, ordered by RA). OW light curves are in g-band and are processed using difference imaging in the OW pipeline. SHOC light curves are observed with no filter, and are binned to match the OW cadence. In each SHOC and OW DFT, the power has been normalised to the peak SHOC or peak OW power resp, for ease of period comparison. See Table~\ref{TAB:validity} for the peak amplitude and associated period for each target.}
\end{figure*}

\begin{figure*}
\centering
\includegraphics[width=\linewidth]{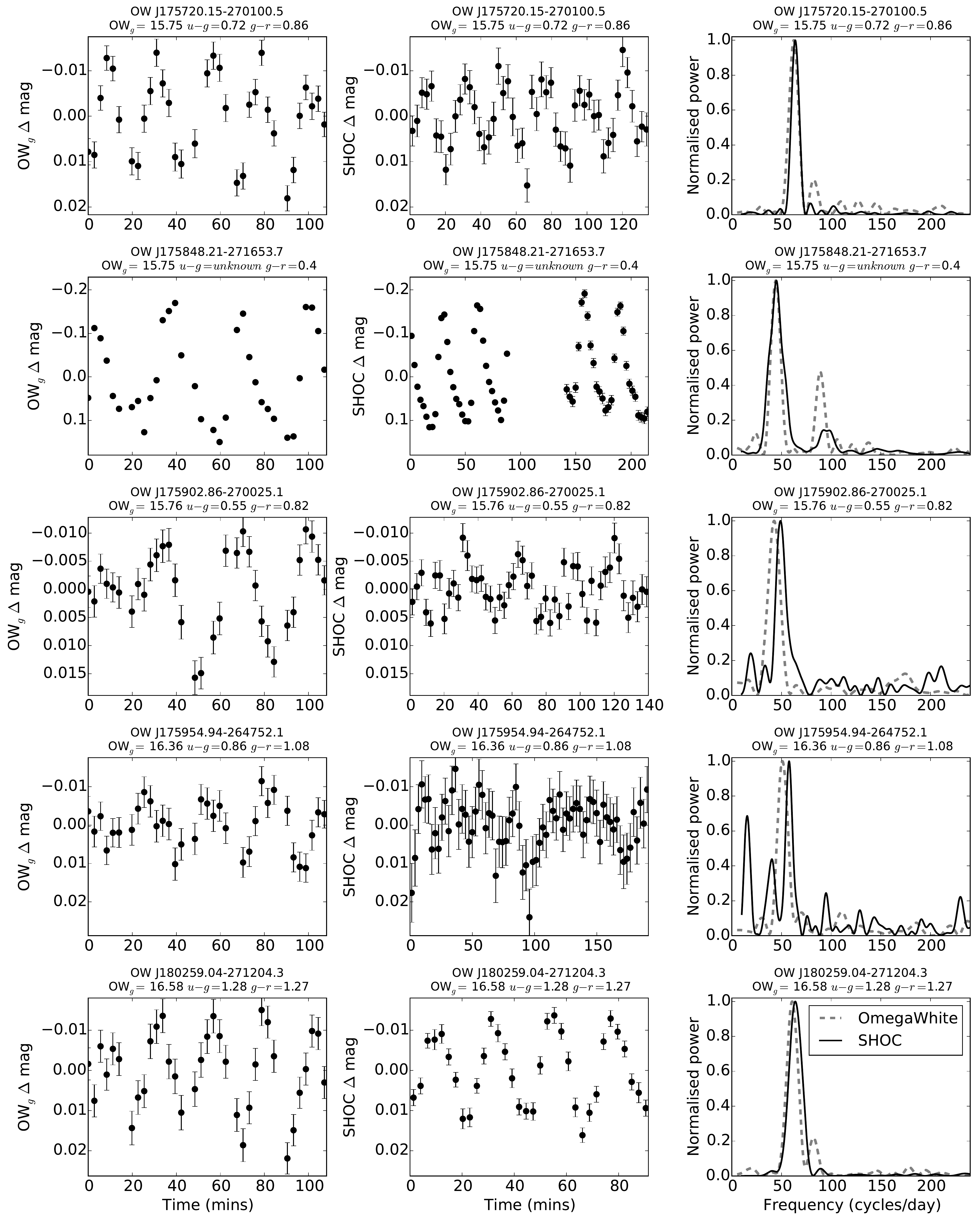}
\contcaption{OW light curves (left panels), follow-up SHOC light curves (middle panels) and discrete Fourier Transform power spectra (`DFT', right panels) of the selected targets (one target per row, ordered by RA). OW light curves are in g-band and are processed using difference imaging in the OW pipeline. SHOC light curves are observed with no filter, and are binned to match the OW cadence. In each SHOC and OW DFT, the power has been normalised to the peak SHOC or peak OW power resp, for ease of period comparison. See Table~\ref{TAB:validity} for the peak amplitude and associated period for each target.}
\end{figure*}

\begin{figure*}
\centering
\includegraphics[width=\linewidth]{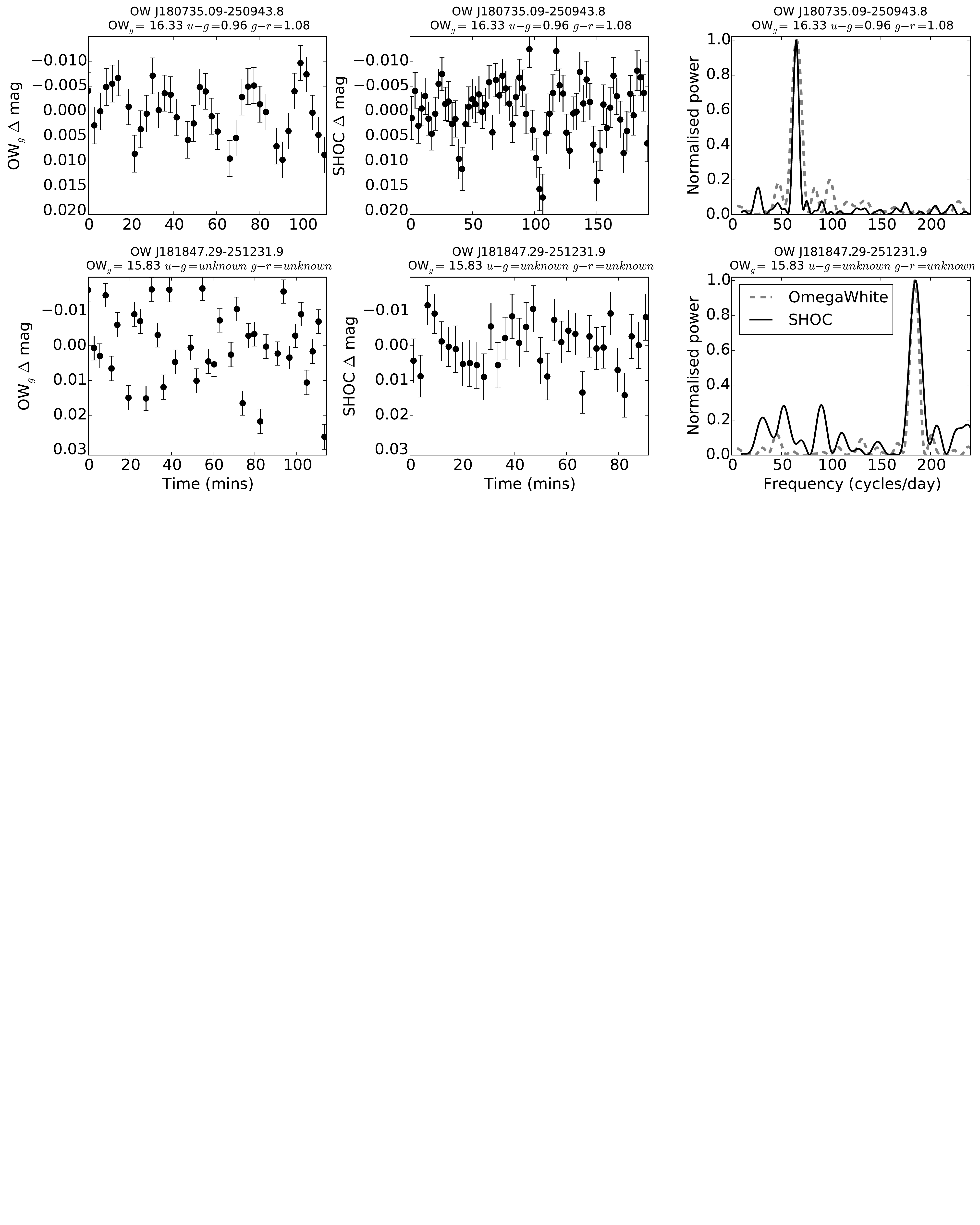}
\contcaption{OW light curves (left panels), follow-up SHOC light curves (middle panels) and discrete Fourier Transform power spectra (`DFT', right panels) of the selected targets (one target per row, ordered by RA). OW light curves are in g-band and are processed using difference imaging in the OW pipeline. SHOC light curves are observed with no filter, and are binned to match the OW cadence. In each SHOC and OW DFT, the power has been normalised to the peak SHOC or peak OW power resp, for ease of period comparison. See Table~\ref{TAB:validity} for the peak amplitude and associated period for each target.}
\end{figure*}

\begin{table*} 
\centering
\caption{Photometric Observing Log for SHOC observations.}
\begin{tabular}{lccccccc}
\hline
Star ID          &  $g_{OW}$   & Date       & Telescope & Duration & Exp. Time & Median Seeing & Std. Dev. Seeing\\
		& (mag)	           	& (dd/mm/yyyy) &           & (hrs) & (s)& (\arcsec)& (\arcsec)\\
\hline
OW$\,$J073919.72--300922.7	&	16.26	&	14/04/2014	&	1.0-m	&	3	&	30	&	1.25	&	0.22	\\
OW$\,$J074106.40--293325.5	&	15.75	&	03/01/2015	&	1.0-m	&	3	&	10	&	1.83	&	0.26	\\
OW$\,$J074447.99--262155.6	&	16.3	&	04/01/2015	&	1.0-m	&	2.7	&	15	&	1.57	&	0.25	\\
OW$\,$J075114.46--270311.2	&	15.69	&	06/01/2015	&	1.0-m	&	3	&	20	&	1.77	&	0.29	\\
OW$\,$J075121.36--270916.4	&	14.63	&	04/01/2015	&	1.0-m	&	5	&	10	&	1.43	&	0.21	\\
OW$\,$J075359.78--275732.6	&	13.98	&	05/01/2015	&	1.0-m	&	3.2	&	10	&	1.14	&	0.29	\\
OW$\,$J075532.58--280854.3	&	16.7	&	31/12/2014	&	1.0-m	&	2.6	&	10	&	1.65	&	0.27	\\
OW$\,$J075837.37--292904.2	&	16.3	&	17/04/2014	&	1.0-m	&	2	&	20	&	1.31	&	0.19	\\
OW$\,$J075952.17--292817.6	&	15.05	&	06/01/2015	&	1.0-m	&	3	&	10	&	2.22	&	0.41	\\
OW$\,$J080205.41--295804.6	&	16.76	&	14/04/2014	&	1.0-m	&	2.7	&	20	&	2.03	&	0.31	\\
OW$\,$J080257.23--292136.0	&	14.71	&	16/04/2014	&	1.0-m	&	3.4	&	30,20	&	2.03&	0.31	\\
OW$\,$J080418.75--293740.2	&	14.17	&	13/04/2014	&	1.0-m	&	1.4	&	10	&	3.09	&	0.33	\\
	&		&	18/04/2014	&	1.0-m	&	2.4	&	10	&	3.83	&	0.63	\\
OW$\,$J080729.17--293417.1	&	17	&	01/01/2015	&	1.0-m	&	3.1	&	30	&	1.94	&	0.28	\\
OW$\,$J080916.72--292158.0	&	15.3	&	03/01/2015	&	1.0-m	&	3.3	&	15	&	2.23	&	0.45	\\
OW$\,$J174308.60--255102.6	&	14.05	&	21/06/2015	&	1.0-m	&	3	&	3	&	1.22	&	0.22	\\
OW$\,$J174436.19--253349.6	&	16.22	&	22/06/2015	&	1.9-m	&	3	&	10	&	1.50	&	0.17	\\
OW$\,$J175122.15--260916.7	&	14.44	&	11/09/2014	&	1.0-m	&	2	&	5	&	1.76	&	0.34	\\
OW$\,$J175210.09--262535.0	&	15.61	&	22/06/2015	&	1.0-m	&	2.7	&	5	&	1.37	&	0.20	\\
OW$\,$J175210.34--262048.7	&	14.69	&	22/06/2015	&	1.9-m	&	3	&	2	&	1.69	&	0.13	\\
OW$\,$J175509.75--260225.5	&	15.31	&	19/06/2015	&	1.9-m	&	1.8	&	5	&	1.74	&	0.33	\\
	&		&	23/06/2015	&	1.9-m	&	3	&	5	&	1.99	&	0.16	\\
OW$\,$J175720.15--270100.5	&	15.75	&	13/09/2014	&	1.0-m	&	3.2	&	10	&	1.54	&	0.30	\\
OW$\,$J175848.21--271653.7	&	15.75	&	20/06/2015	&	1.9-m	&	2.7	&	15,10	&	2.63	&	0.91	\\
OW$\,$J175902.86--270025.1	&	15.76	&	21/06/2015	&	1.9-m	&	2.3	&	10	&	1.54	&	0.13	\\
OW$\,$J175954.94--264752.1	&	16.36	&	15/09/2014	&	1.0-m	&	3.2	&	5	&	1.28	&	0.14	\\
OW$\,$J180259.04--271204.3	&	16.58	&	19/06/2015	&	1.9-m	&	1.5	&	10	&	2.04	&	0.15	\\
	&		&	20/06/2015	&	1.9-m	&	1.5	&	10	&	1.58	&	0.12	\\
	&		&	21/06/2015	&	1.9-m	&	1.5	&	10	&	2.06	&	0.19	\\
OW$\,$J180735.09--250943.8	&	16.33	&	12/09/2014	&	1.0-m	&	3.2	&	10	&	1.56	&	0.42	\\
OW$\,$J181847.29--251231.9	&	15.83	&	10/09/2014	&	1.0-m	&	3	&	5	&	1.12	&	0.23	\\

\hline
\end{tabular}
\label{TAB:obslog_phot}
\end{table*}

\begin{table*} 
\centering
\caption{Spectroscopic Observing Log for SALT and SpUpNIC (1.9-m) observations} 
\begin{tabular}{lccccccc}
\hline
Star ID      &  $g_{OW}$           &  Obs. Date        & Telescope & Grating & Gr. Angle & Slit Width& Exp. Time        \\
                & (mag)	         & (dd/mm/yyyy) &           &         & $\degr$ & $\arcsec$  & (s)               \\
\hline
OW$\,$J073649.27--295601.8 & 14.66 & 08/01/2016 & 1.9-m & gr7    & 16.7   & 1.95 & 1200              \\
OW$\,$J073823.16--303958.0 & 15.70 & 11/01/2016 & 1.9-m & gr7    & 16.65  & 2.39 & 1200              \\
OW$\,$J073909.96--300620.0 & 13.64 & 19/12/2015 & 1.9-m & gr6    & 12.2   & 1.8  & 900               \\
OW$\,$J074059.42--242452.4 & 14.01 & 11/01/2016 & 1.9-m & gr7    & 16.8   & 1.8  & 900               \\
OW$\,$J074101.18--291540.0 & 16.65 & 18/12/2015 & 1.9-m & gr7    & 17     & 1.35 & 1200              \\
OW$\,$J074106.07--294811.0 & 20.02 & 13/04/2015 & SALT  & pg0300 & 5      & 1.5  & 1000 ($\times$ 2) \\
                       &       & 23/04/2015 & SALT  & pg0300 & 5      & 1.5  & 1000 ($\times$ 2) \\
OW$\,$J074106.40--293325.5 & 15.75 & 30/12/2014 & SALT  & pg0300 & 5      & 1.5  & 180 ($\times$ 3)  \\
                       &       & 01/01/2016 & 1.9-m & gr7    & 16.65  & 2.69 & 1500              \\
OW$\,$J074127.67--243642.9 & 15.39 & 12/01/2016 & 1.9-m & gr7    & 16.9   & 1.8  & 1500              \\
OW$\,$J074216.77--241327.5 & 15.99 & 11/01/2016 & 1.9-m & gr7    & 16.7   & 2.1  & 1500              \\
OW$\,$J074447.99--262155.6 & 16.30 & 09/01/2016 & 1.9-m & gr7    & 16.78  & 1.8  & 1500              \\
OW$\,$J074513.22--261036.0 & 14.73 & 11/01/2015 & SALT  & pg0300 & 5      & 1.5  & 180 ($\times$ 3)  \\
                       &       & 10/01/2016 & 1.9-m & gr7    & 16.65  & 2.69 & 1500              \\
OW$\,$J074515.73--260841.7 & 15.11 & 19/12/2015 & 1.9-m & gr6    & 12.2   & 1.95 & 1200              \\
OW$\,$J074533.75--263212.5 & 15.16 & 20/12/2015 & 1.9-m & gr6    & 12.2   & 2.54 & 1200              \\
OW$\,$J074551.57--242147.8 & 15.71 & 12/01/2016 & 1.9-m & gr7    & 16.9   & 1.8  & 1500              \\
OW$\,$J075049.73--270018.3 & 16.03 & 21/12/2015 & 1.9-m & gr6    & 12.6   & 1.5  & 1200              \\
OW$\,$J075114.46--270311.2 & 15.69 & 22/12/2015 & 1.9-m & gr6    & 12.6   & 1.5  & 900               \\
OW$\,$J075121.36--270916.4 & 14.63 & 10/01/2016 & 1.9-m & gr7    & 16.7   & 1.8  & 900               \\
OW$\,$J075305.78--301208.1 & 16.93 & 28/11/2015 & SALT  & pg0900 & 14.375 & 1.5  & 700               \\
OW$\,$J075359.78--275732.6 & 13.98 & 27/12/2014 & SALT  & pg0300 & 5      & 1.5  & 180 ($\times$ 3)  \\
                       &       & 10/01/2016 & 1.9-m & gr7    & 16.75  & 1.5  & 600               \\
OW$\,$J075436.40--304655.0 & 20.62 & 01/01/2016 & SALT  & pg0900 & 14.375 & 1.5  & 2000              \\
OW$\,$J075527.61--314825.3 & 15.85 & 04/01/2016 & SALT  & pg0900 & 14.375 & 1.5  & 600               \\
                       &       & 06/01/2016 & 1.9-m & gr7    & 16.5   & 2.39 & 1200              \\
OW$\,$J075531.59--315058.2 & 16.98 & 08/01/2016 & 1.9-m & gr7    & 16.7   & 2.1  & 1500              \\
OW$\,$J075532.58--280854.3 & 16.70 & 11/01/2015 & SALT  & pg0300 & 5      & 1.5  & 180 ($\times$ 3)  \\
OW$\,$J075620.67--291605.4 & 16.95 & 08/01/2016 & 1.9-m & gr7    & 16.7   & 2.1  & 1500              \\
OW$\,$J075640.83--291107.9 & 15.83 & 12/01/2016 & 1.9-m & gr7    & 16.87  & 1.5  & 1200              \\
OW$\,$J075719.94--292955.5 & 20.24 & 02/01/2016 & SALT  & pg0900 & 14.375 & 1.5  & 2000              \\
OW$\,$J075837.37--292904.2 & 16.30 & 09/01/2016 & 1.9-m & gr7    & 16.65  & 1.5  & 1500              \\
OW$\,$J075918.25--313721.6 & 15.73 & 12/01/2016 & 1.9-m & gr7    & 16.75  & 1.8  & 1200              \\
OW$\,$J075952.17--292817.6 & 15.05 & 09/01/2016 & 1.9-m & gr7    & 16.7   & 2.69 & 1200              \\
OW$\,$J080146.04--294518.2 & 15.93 & 10/01/2016 & 1.9-m & gr7    & 16.65  & 2.69 & 1500              \\
OW$\,$J080154.61--281513.9 & 17.11 & 09/01/2016 & 1.9-m & gr7    & 16.65  & 1.5  & 1500              \\
OW$\,$J080202.79--302356.3 & 15.81 & 09/01/2016 & 1.9-m & gr7    & 16.7   & 2.69 & 1500              \\
OW$\,$J080205.31--300331.4 & 14.54 & 22/12/2015 & 1.9-m & gr6    & 12.6   & 1.5  & 900               \\
OW$\,$J080205.41--295804.6 & 16.76 & 09/01/2016 & 1.9-m & gr7    & 16.65  & 1.5  & 1500              \\
OW$\,$J080242.91--314653.4 & 15.72 & 12/01/2016 & 1.9-m & gr7    & 16.75  & 2.39 & 1200              \\
OW$\,$J080257.23--292136.0 & 14.71 & 11/01/2016 & 1.9-m & gr7    & 16.7   & 2.1  & 900               \\
OW$\,$J080311.83--291144.6 & 19.74 & 08/12/2015 & SALT  & pg0900 & 14.375 & 1.5  & 2000              \\
OW$\,$J080313.04--320720.2 & 16.80 & 11/01/2016 & 1.9-m & gr7    & 16.7   & 2.69 & 1500              \\
OW$\,$J080323.23--314852.6 & 16.09 & 12/01/2016 & 1.9-m & gr7    & 16.8   & 2.1  & 1200              \\
OW$\,$J080418.75--293740.2 & 14.17 & 09/01/2016 & 1.9-m & gr7    & 16.7   & 2.69 & 1200              \\
OW$\,$J080517.37--310257.5 & 15.55 & 11/01/2016 & 1.9-m & gr7    & 16.7   & 2.69 & 1200              \\
OW$\,$J080527.94--275144.8 & 15.52 & 22/12/2015 & 1.9-m & gr6    & 12.6   & 1.5  & 1200              \\
OW$\,$J080531.16--274221.5 & 17.59 & 19/12/2015 & 1.9-m & gr7    & 17     & 1.35 & 1200              \\
OW$\,$J080626.33--302545.3 & 15.68 & 08/01/2016 & 1.9-m & gr7    & 16.7   & 2.1  & 1200              \\
OW$\,$J080901.66--291652.2 & 16.24 & 12/01/2016 & 1.9-m & gr7    & 16.75  & 2.39 & 1500              \\
OW$\,$J080910.70--300744.1 & 16.19 & 09/01/2016 & 1.9-m & gr7    & 16.65  & 1.8  & 1500              \\
OW$\,$J080911.11--300209.2 & 15.96 & 10/01/2016 & 1.9-m & gr7    & 16.65  & 2.69 & 1500              \\
OW$\,$J080916.72--292158.0 & 15.30 & 30/12/2014 & SALT  & pg0300 & 5      & 1.5  & 180 ($\times$ 3)  \\
                       &       & 11/01/2016 & 1.9-m & gr7    & 16.65  & 2.99 & 1500              \\
OW$\,$J081018.30--273749.1 & 15.49 & 20/12/2015 & 1.9-m & gr6    & 12.6   & 1.5  & 1200              \\
OW$\,$J081116.72--313821.1 & 14.83 & 12/01/2016 & 1.9-m & gr7    & 16.87  & 1.5  & 900               \\
OW$\,$J081234.73--312409.8 & 15.02 & 12/01/2016 & 1.9-m & gr7    & 16.8   & 1.8  & 1200              \\
OW$\,$J081321.93--313537.7 & 14.86 & 12/01/2016 & 1.9-m & gr7    & 16.75  & 1.8  & 900               \\
OW$\,$J082257.05--280051.0 & 15.52 & 12/01/2016 & 1.9-m & gr7    & 16.85  & 1.5  & 1200              \\
OW$\,$J172235.29--321403.6 & 15.75 & 27/04/2016 & 1.9-m & gr7    & 16.0 & 1.5 & 1200/1500        \\
OW$\,$J175848.21--271653.7 & 15.75 & 01/05/2016 & 1.9-m & gr7    & 16.0 & 1.8 & 1800/1500        \\
OW$\,$J180441.23--270912.4 & 15.06 & 27/04/2016 & 1.9-m & gr7    & 16.0   & 1.5  & 1200 ($\times$ 3) \\
OW$\,$J181043.86--275750.0 & 14.70 & 27/04/2016 & 1.9-m & gr7    & 16.0 & 1.5 & 1200 ($\times$ 3)        \\
\hline
\end{tabular}
\label{obslog_spec}
\end{table*}

\pagebreak

\begin{sidewaysfigure*}
\centering
\includegraphics[width=\linewidth]{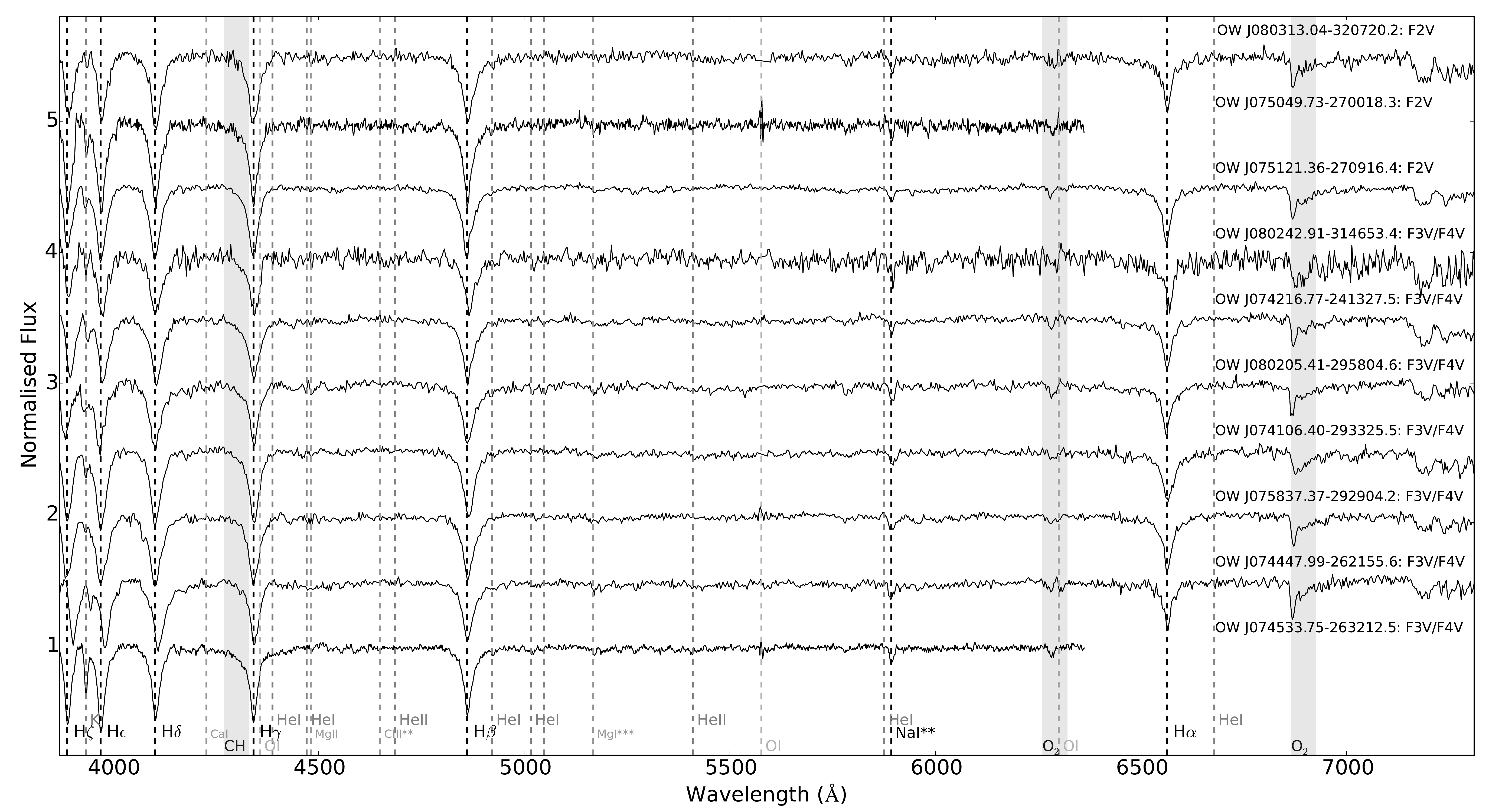}
\caption{Continuum-normalised spectra of the SpUpNIC-observed targets, ordered by proposed spectral type (not including the three SIMBAD targets presented in Section~\ref{SIMBAD}). An offset to each spectrum has been applied for visualisation purposes. Dashed lines indicate important spectral lines (as labeled), and shaded areas cover either atmospheric telluric bands or the G-band (CH).}
\label{FIG:allspec_1}
\end{sidewaysfigure*}

\begin{sidewaysfigure*}
\centering
\includegraphics[width=\linewidth]{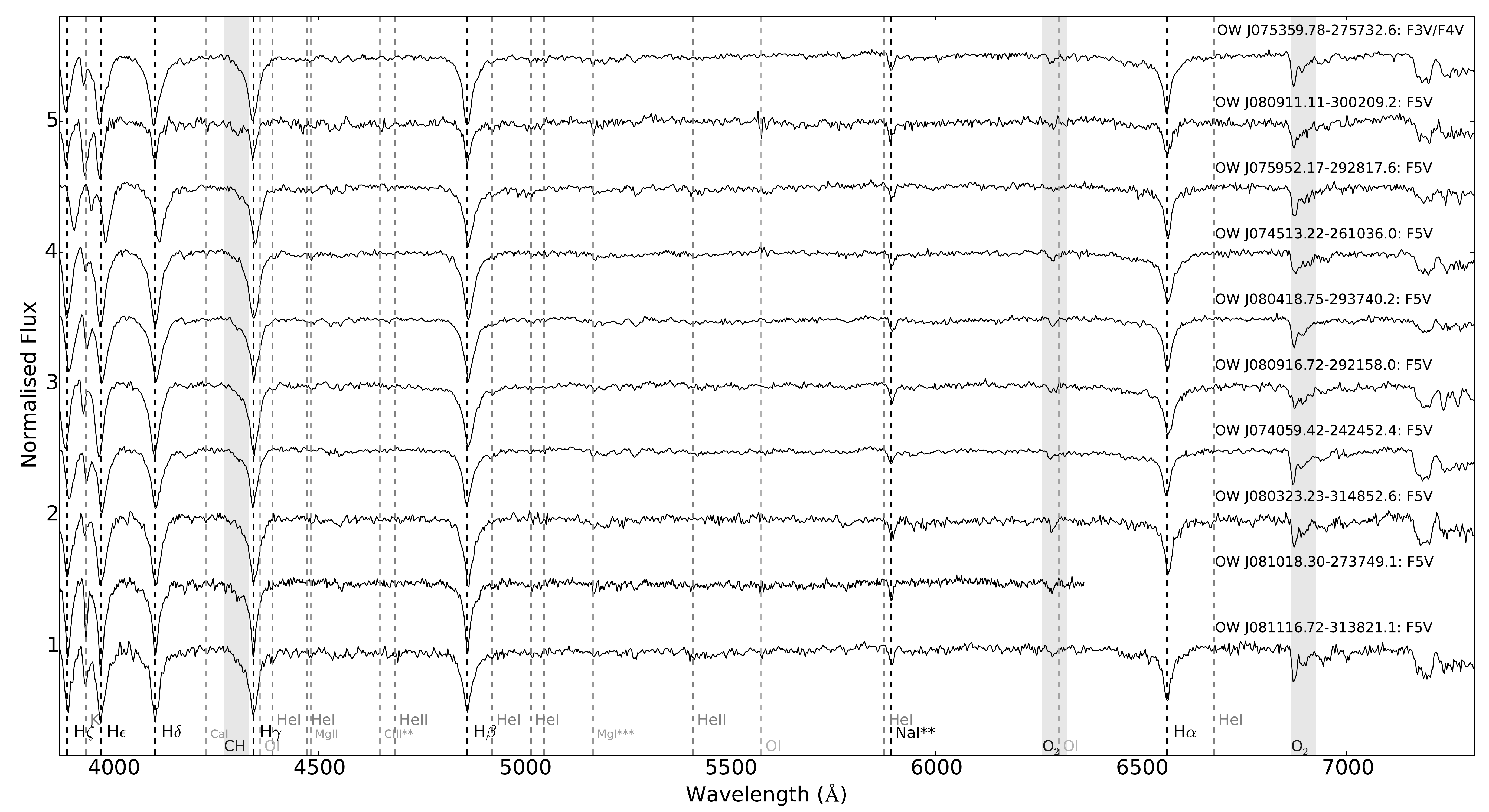}
\contcaption{Continuum-normalised spectra of the SpUpNIC-observed targets, ordered by proposed spectral type (not including the three SIMBAD targets presented in Section~\ref{SIMBAD}). An offset to each spectrum has been applied for visualisation purposes. Dashed lines indicate important spectral lines (as labeled), and shaded areas cover either atmospheric telluric bands or the G-band (CH).}
\end{sidewaysfigure*}

\begin{sidewaysfigure*}
\centering
\includegraphics[width=\linewidth]{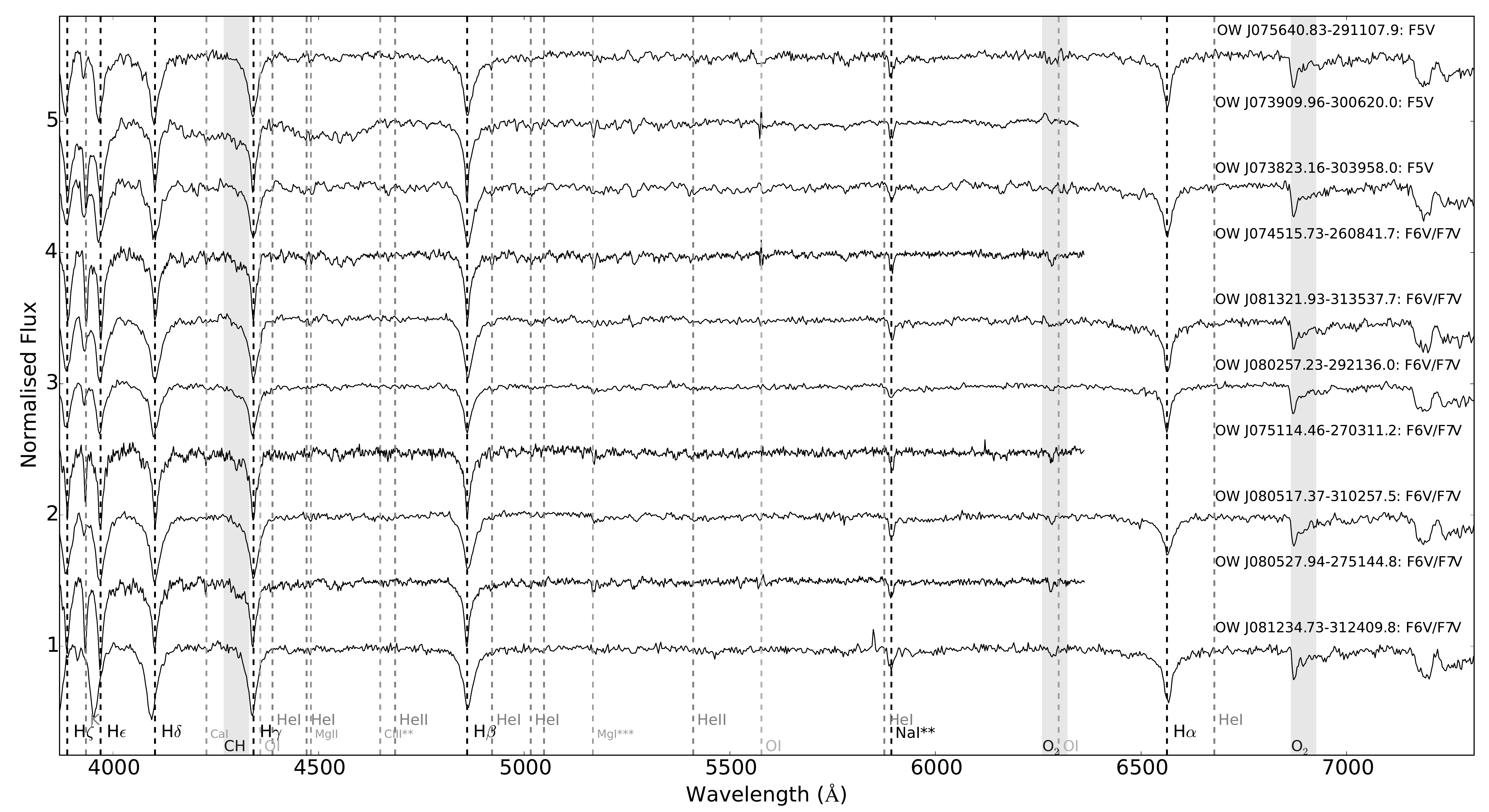}
\contcaption{Continuum-normalised spectra of the SpUpNIC-observed targets, ordered by proposed spectral type (not including the three SIMBAD targets presented in Section~\ref{SIMBAD}). An offset to each spectrum has been applied for visualisation purposes. Dashed lines indicate important spectral lines (as labeled), and shaded areas cover either atmospheric telluric bands or the G-band (CH).}
\end{sidewaysfigure*}

\begin{sidewaysfigure*}
\centering
\includegraphics[width=\linewidth]{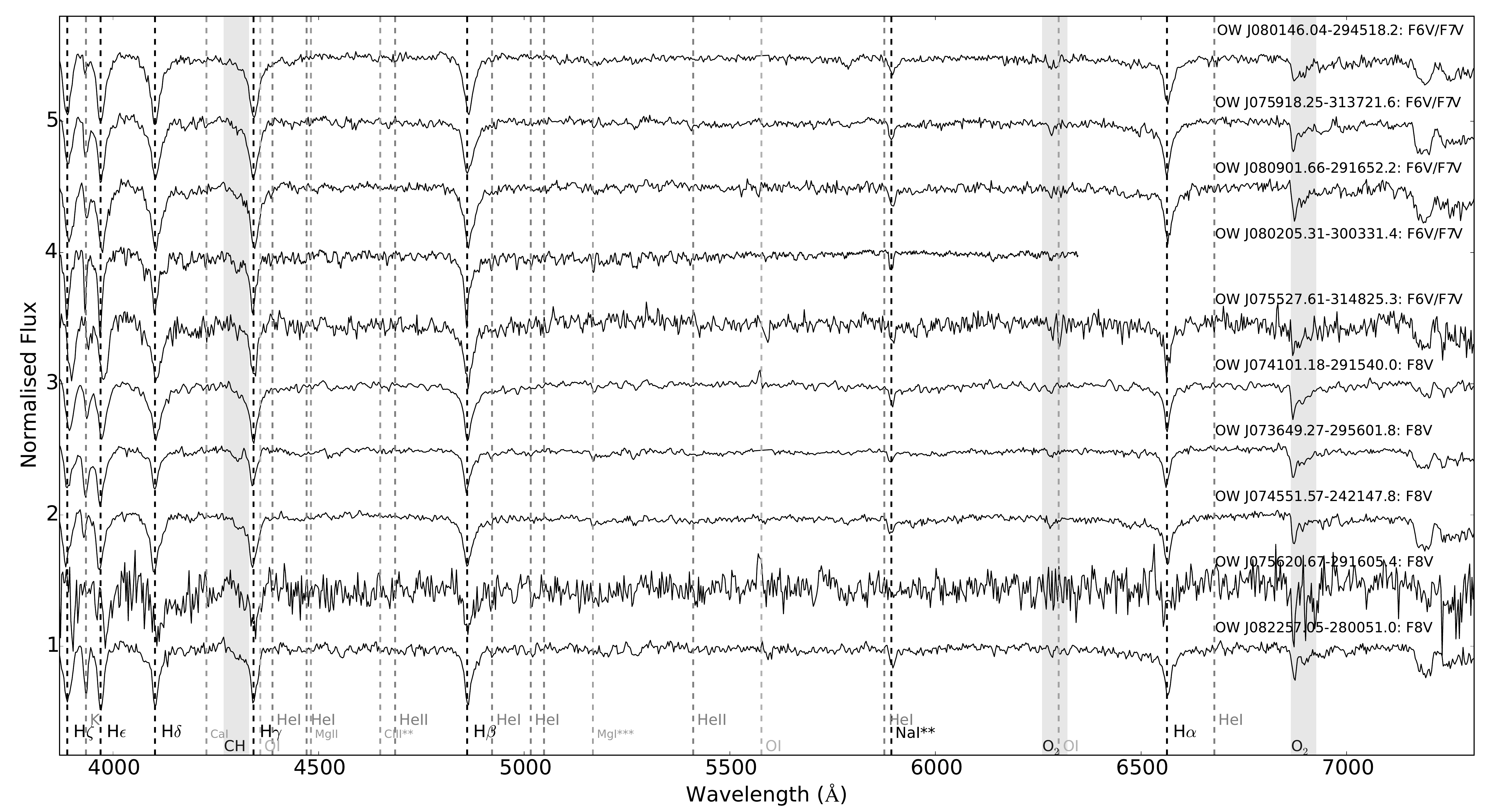}
\contcaption{Continuum-normalised spectra of the SpUpNIC-observed targets, ordered by proposed spectral type (not including the three SIMBAD targets presented in Section~\ref{SIMBAD}). An offset to each spectrum has been applied for visualisation purposes. Dashed lines indicate important spectral lines (as labeled), and shaded areas cover either atmospheric telluric bands or the G-band (CH).}
\end{sidewaysfigure*}

\begin{sidewaysfigure*}
\centering
\includegraphics[width=\linewidth]{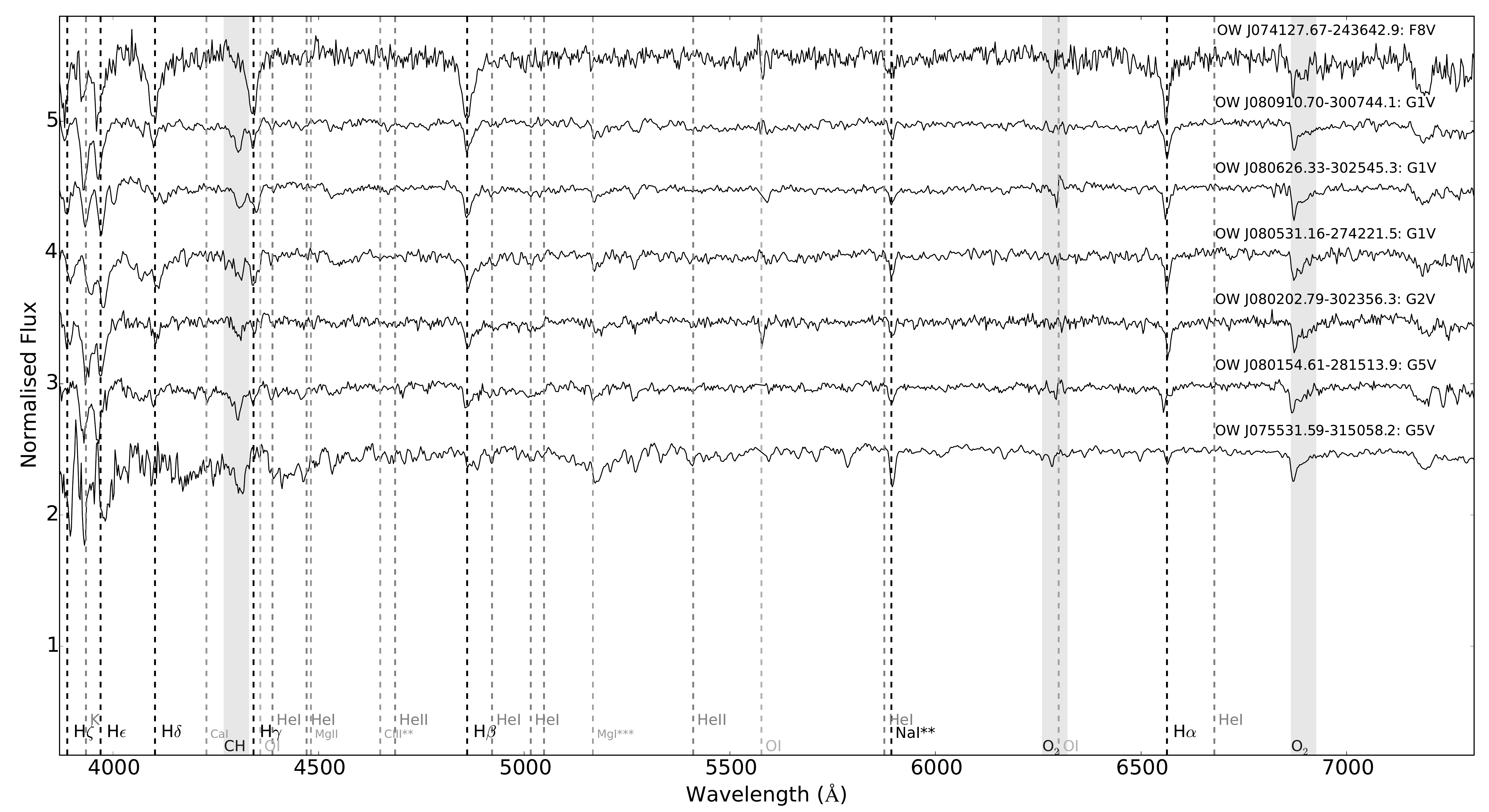}
\contcaption{Continuum-normalised spectra of the SpUpNIC-observed targets, ordered by proposed spectral type (not including the three SIMBAD targets presented in Section~\ref{SIMBAD}). An offset to each spectrum has been applied for visualisation purposes. Dashed lines indicate important spectral lines (as labeled), and shaded areas cover either atmospheric telluric bands or the G-band (CH).}
\end{sidewaysfigure*}

\bsp	
\label{lastpage}
\end{document}